\documentclass{aa}  

\usepackage{gensymb}

\newcommand*\eup{$E_{\rm up}$}
\newcommand{\lsun}{\mbox{L}_\odot}
\newcommand{\msun}{\mbox{M}_\odot}

\usepackage{graphicx}
\usepackage{txfonts}
\begin{document} 

    \title{JWST Observations of Young protoStars (JOYS)}
    \titlerunning{JOYS-TMC1 }
    \authorrunning{Tychoniec et al. }

     \subtitle{Linked accretion and ejection in a Class I protobinary system}

%   \subtitle{I. Overviewing the $\kappa$-mechanism}

   \author{Łukasz Tychoniec \inst{1,2}, Martijn L. van Gelder \inst{1}, Ewine F. van Dishoeck \inst{1,3}, Logan Francis \inst{1}, Will R. M. Rocha \inst{1,4}, \\ Alessio Caratti o Garatti \inst{5}, Henrik Beuther \inst{6}, Caroline Gieser \inst{3},  Kay Justtanont \inst{7}, Harold Linnartz \thanks{Deceased on December 31, 2023}\inst{1,4}, 
\\ 
Valentin J.~M. Le Gouellec \inst{8,9}, Giulia Perotti \inst{6}, R. Devaraj \inst{10}, Benoît Tabone \inst{11}, Thomas P. Ray \inst{10},\\ Nashanty G. C. Brunken \inst{1}, Yuan Chen \inst{1}, Patrick J. Kavanagh \inst{12}, Pamela Klaassen\inst{13}, Katerina Slavicinska \inst{1,4},\\ Manuel Güdel \inst{6,14,15}, Goran Östlin \inst{16}. 
  }
   \institute{Leiden Observatory, Leiden University, PO Box 9513, 2300RA, Leiden, The Netherlands\\
                 \email{tychoniec@strw.leidenuniv.nl}
                 \and
European Southern Observatory, Karl-Schwarzschild-Strasse 2, 85748 Garching bei M\"unchen, Germany
\and
Max-Planck-Institut f{\"u}r Extraterrestrische Physik, Giessenbachstrasse 1, D-85748 Garching, Germany
\and 
Laboratory for Astrophysics, Leiden Observatory, Leiden University, PO Box 9513, NL 2300 RA Leiden, The Netherlands
\and 
INAF-Osservatorio Astronomico di Capodimonte, Salita Moiariello 16, I-80131 Napoli, Italy
\and 
Max Planck Institute for Astronomy, Königstuhl 17, 69117 Heidelberg, Germany
\and 
Department of Space, Earth and Environment, Chalmers University of Technology, Onsala Space Observatory, 439 92 Onsala, Sweden
\and
NASA Ames Research Center, Space Science and Astrobiology Division M.S.245-6 Moffett Field, CA94035, USA 
\and
NASA Postdoctoral Program Fellow
\and
School of Cosmic Physics, Dublin Institute for Advanced Studies, 31 Fitzwilliam Place, Dublin 2, Ireland
\and
Université Paris-Saclay, CNRS, Institut d’Astrophysique Spatiale, 91405, Orsay,France
\and
Department of Experimental Physics, Maynooth University, Maynooth, Co Kildare, Ireland
\and
 UK Astronomy Technology Centre, Royal Observatory Edinburgh, Blackford Hill, Edinburgh EH9 3HJ, UK
 \and
 Dept. of Astrophysics, University of Vienna, Türkenschanzstr. 17, A-1180 Vienna, Austria
 \and
ETH Zürich, Institute for Particle Physics and Astrophysics, Wolfgang-Pauli-Str. 27, 8093 Zürich, Switzerland
\and
Department of Astronomy, Oskar Klein Centre, Stockholm University, 106 91 Stockholm, Sweden
}

   \date{Received: December 11, 2023}

% \abstract{}{}{}{}{} 
% 5 {} token are mandatory
 
  \abstract
  % context heading (optional)
  % {} leave it empty if necessary  
   {Accretion and ejection dictate the outcomes of star and planet formation processes. The mid-infrared (MIR) wavelength range offers key tracers of  processes that have been difficult to detect and spatially resolve in protostars until now. }
  % aims heading (mandatory)
   {We aim to characterize the interplay between accretion and ejection in the low-mass Class I protobinary system TMC1, comprising two young stellar objects: TMC1-W and TMC1-E at a 85 au separation.}
  % methods heading (mandatory)
   {Using the {\it James Webb} Space Telescope (JWST) Mid-Infrared Instrument (MIRI) observations in 5 -- 28 $\mu$m range, we measured the intensities of emission lines of H$_2$, atoms, and ions, for instance, the\ [Fe II] and [Ne II], and HI recombination lines. We analyzed the spatial distribution of the different species using the MIRI Medium Resolution Spectrometer (MRS) capabilities to spatially resolve emission at 0\farcs2--0\farcs7 scales. We compared these results with the corresponding Atacama Large Millimeter/submillimeter Array (ALMA) maps tracing cold gas and dust.}
  % results heading (mandatory)
   {We detected H$_2$ outflow coming from TMC1-E, with no significant H$_2$ emission from TMC1-W. The H$_2$ emission from TMC1-E outflow appears narrow and extends to wider opening angles with decreasing \eup\ from S(8) to S(1) rotational transitions, indicating the disk wind as its origin. The outflow from TMC1-E protostar shows spatially extended emission lines of [Ne II], [Ne III], [Ar II], and [Ar III], with their line ratios consistent with UV radiation as a source of ionization. With ALMA, we detected an accretion streamer infalling from $>$ 1000 au scaling onto the TMC1-E component. The TMC1-W protostar powers a collimated jet, detected with [Fe II] and [Ni II], making it consistent with energetic flow. A much weaker ionized jet is observed from TMC1-E, and both jets appear strikingly parallel to each other, indicating that the disks are co-planar. TMC1-W is associated with strong emission from hydrogen recombination lines, tracing the accretion onto the young star.}
  % conclusions heading (optional), leave it empty if necessary 
   {MIRI-MRS observations provide an unprecedented view of protostellar accretion and ejection processes on 20 au scales. Observations of a binary Class I protostellar system show that the two processes are clearly intertwined, with accretion from the envelope onto the disk influencing a wide-angle wind ejected on disk scales. Finally, the accretion from the protostellar disk onto the protostar is associated with the source launching a collimated high-velocity jet within the innermost regions of the disk.
}

   \keywords{}

   \maketitle
%
%-------------------------------------------------------------------

\section{Introduction}
The interplay between the accretion and ejection of mass sets the outcome of the star formation process \citep{Hartmann.Herczeg.ea2016}. However, the link between those two key processes is difficult to ascertain. It is expected that the excess of angular momentum is lost during the accretion from the envelope to the disk or from the disk to the protostar and is removed by the outflow of material either by the jet (collimated, high-velocity stream of gas) or wide-angle disk wind (with a broader opening angle and lower velocities) For more details 
we refer to \cite{Frank.Ray.ea2014, Pascucci.Cabrit.ea2022} and references therein.
The emerging paradigm of the protostellar outflow launching is that the outflow is magnetohydrodynamically (MHD) driven \citep{Pascucci.Cabrit.ea2022}, with the outflow launching point and radius still a matter of debate. Observational evidence shows wide-angle winds launched from the disk \citep{Bjerkeli.vanderWiel.ea2016, Louvet.Dougados.ea2018} and collimated jets launched near or within the dust sublimation radius \citep{Tabone.Cabrit.ea2017,Lee.Codella.ea2019}. However, the balance and interaction between the jets and disk winds have not been well characterized yet.

Binary systems are unique laboratories for studying protostellar accretion and ejection since it is possible to compare the properties of the two protostars directly. Moreover, interactions between binary members may have strong effects on the truncation of the disk and triggering of accretion outbursts. Since many low-mass protostars are born in binary systems, capturing their formation and evolution is important for understanding the star formation process. \citep{Raghavan.McAlister.ea2010, Tobin.Looney.ea2016, Tobin.Sheehan.ea2020, Offner.Moe.ea2023}. However, such studies in the mid-infrared (MIR) regime are scarce, limited by the spatial resolution of instruments offered at these wavelengths, while many essential tracers of outflow activity (e.g., H$_2$, [Fe II]) are found in this wavelength range. This regime is crucial, especially for the studies of the youngest protostars, where extinction limits the possibility of studying those sources in optical and near-IR. 

Infrared emission lines have been studied in protostars and their outflows, especially with {\it Spitzer} Infrared Spectrograph (IRS) spectroscopy with a typical aperture of 5\arcsec 
 \citep[e.g.,][]{Maret.Bergin.ea2009,Lahuis.vanDishoeck.ea2010, Tappe.Forbrich.ea2012}.
{\it Spitzer}-IRS observations already showed hints that binary systems might present very different outflow properties such as H$_2$ lines dominating emission from one source and ion lines from the other \citep{Dionatos.Joergensen.ea2014}. However, it has been challenging to map the protostellar outflows and jets in the MIR at sub-arcsecond scales without the integrated field unit (IFU) capabilities to retrieve spectral information from each spatial element on sub-arcsecond scales.

Accretion onto the protostar and infall from envelope to disk are typically probed in very different energy regimes, with stellar accretion primarily studied with hydrogen recombination lines \citep{Alcala.Manara.ea2017}, while infall is probed by kinematic features in the cool gas in the sub-millimeter regime \citep{Joergensen.Schoeier.ea2002, Aso.Ohashi.ea2015}. Therefore, the combination of the two, augmented by MIR diagnostics, can give us a complete understanding of the transfer of the material all the way from the large-scale envelope through the young disk onto the protostar. 

TMC1, also known as IRAS 04381+2540, is a nearby Class I protobinary star system located in the Taurus-L1527 region \citep[141.8$\pm$1.4 pc;][]{Krolikowski.Kraus.ea2021}. The system contains two protostars associated with near-IR emission tracing scattered light from the outflow cavity walls and the near-IR jet. \citep{Apai.Toth.ea2005, Terebey.VanBuren.ea2006}.  When observed with Atacama Large Millimeter/submillimeter Array (ALMA) at sub-arcsecond resolution at 1.3 mm, the binary stars were resolved with the projected separation of 85 au \citep{vantHoff.Harsono.ea2020}. 

In this paper, we present observations of the TMC1 protobinary system with Medium Resolution Spectroscopy mode \citep[MRS;][]{Wells.Pel.ea2015, Argyriou.Glasse.ea2023} of the Mid-InfraRed Instrument \citep[MIRI;][]{Wright.Wright.ea2015, Rieke.Ressler.ea2015, Wright.Rieke.ea2023} on board the {\it James Webb} Space Telescope \citep[JWST;][]{Rigby.Perrin.ea2023}. Complementary ALMA observations already presented in \cite{Tychoniec.vanDishoeck.ea2021} are used to compare the MIR emission with cold gas tracers.

This paper presents stunning differences in the appearance of outflow in the TMC1 binary system. This work is organized as follows. In Section 2, we introduce JWST/MIRI MRS and ALMA observations; Section 3 describes the results, including spectral and spatial analysis of the system; Section 4 discusses the origin of differences in this system and Section 5 provides our conclusions.

\begin{figure*}
\begin{center}
    \includegraphics[width=0.95\textwidth]{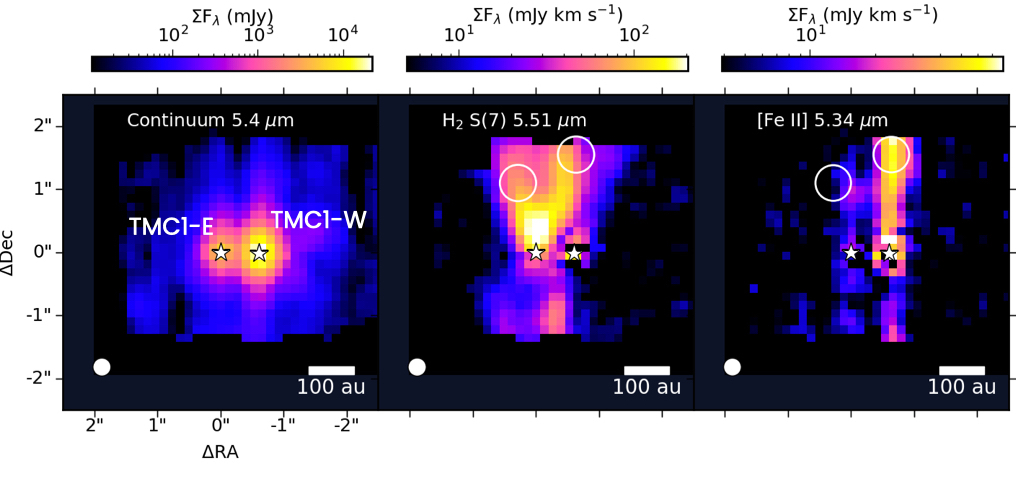}
    \caption{ {\it } Continuum emission at 5.4 $\mu$m obtained from line-free spectral channels in Channel 1 (left). {\it Center:} Integrated emission of H$_2$ 0-0 S(7) line at 5.51 $\mu$m in colorscale. {\it Right:} Integrated emission of [Fe II] line at 5.34 $\mu$m in color scale.  Stars mark the MIR peak continuum position of the protostars. White circles show regions from which the spectra at the outflow positions were obtained, with diameters corresponding to two times the MIRI-MRS point-spread function (PSF) at 5.5  $\mu$m. In the bottom-left corners, MIRI-MRS empirical FWHM of PSF \citep{Law.Morrison.ea2023} is indicated as a white circle.}
    %\caption{Integrated emission of a line-free part of the spectra in each of the MIRI-MRS channels. The colorscale spans from maximum flux down to 0.05$\%$ of the maximum value. In each channel, the regions are overlaid from which the spectra for the analysis were extracted: 1 - blueshifted wind from TMC1-E; 2 - blueshifted jet from TMC1-W; 3 - TMC1-E source position; 4 - TMC1-W source position; 5 - TMC1-E redshifted wind; 6 - TMC1-W redshifted jet. } 
    \label{fig:representative_regs}
    %Contsub_Chanmaps_Final Figure?

    %CHANGES: this needs to be updated for the final regions.
\end{center}
\end{figure*}

\begin{figure*}
\begin{center}
    \includegraphics[width=0.95\textwidth]{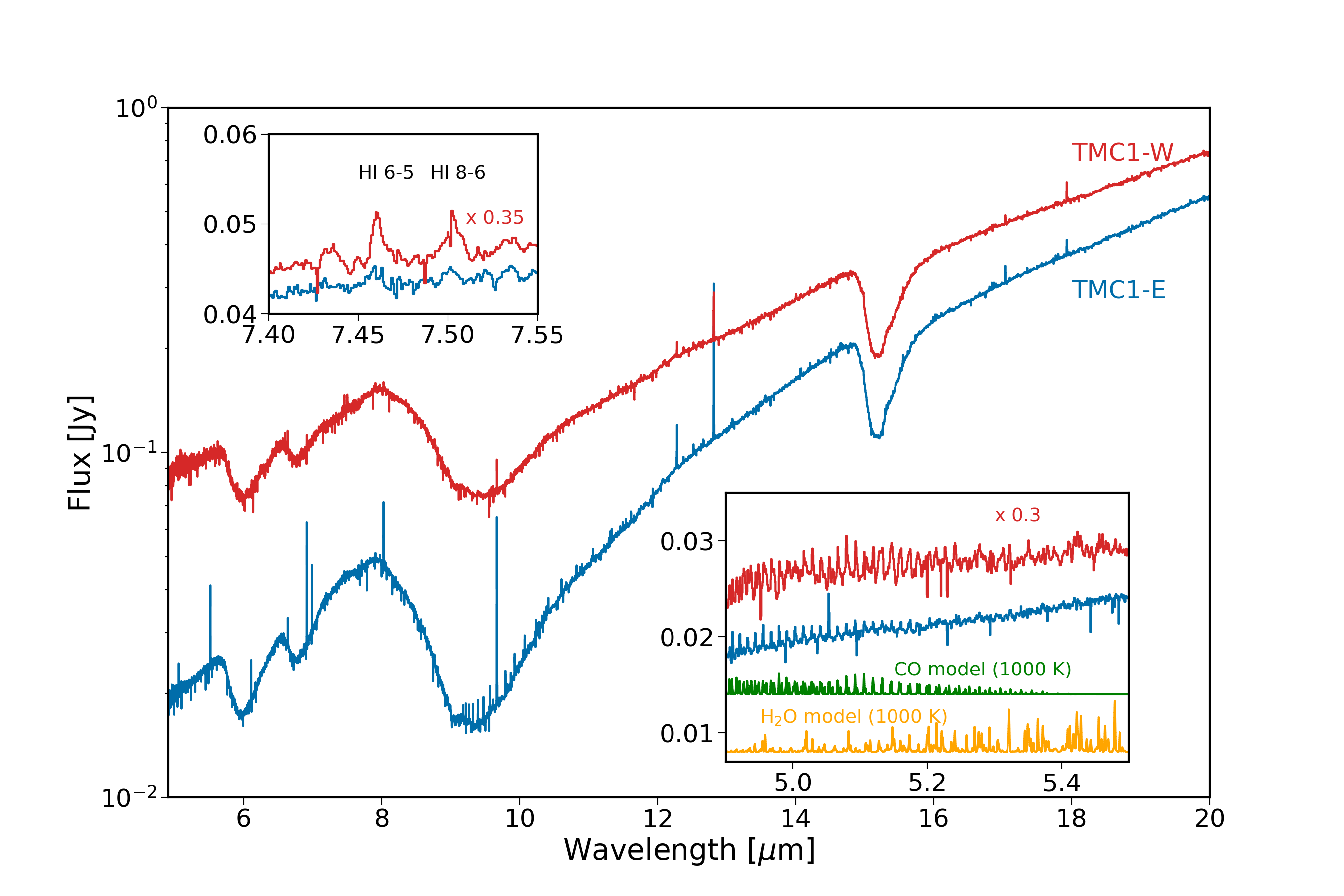}

    \caption{Spectra extracted from the MIR continuum peaks of the two protostars. Several absorption features associated with ice and dust are seen in both sources and emission lines coming from the disk and central protostars, although some contamination with outflows and winds is also possible. {\it Top-left:} Zoom-in on the two hydrogen recombination lines detected in emission towards protostars. {\it Bottom-right: } Spectra revealing molecular gas emission and corresponding slab models of CO (green) and H$_2$O (orange). }
    \label{fig:spectra_overview}
\end{center}
\end{figure*}

\begin{figure*}
\begin{center}
        \includegraphics[width=0.95\textwidth]{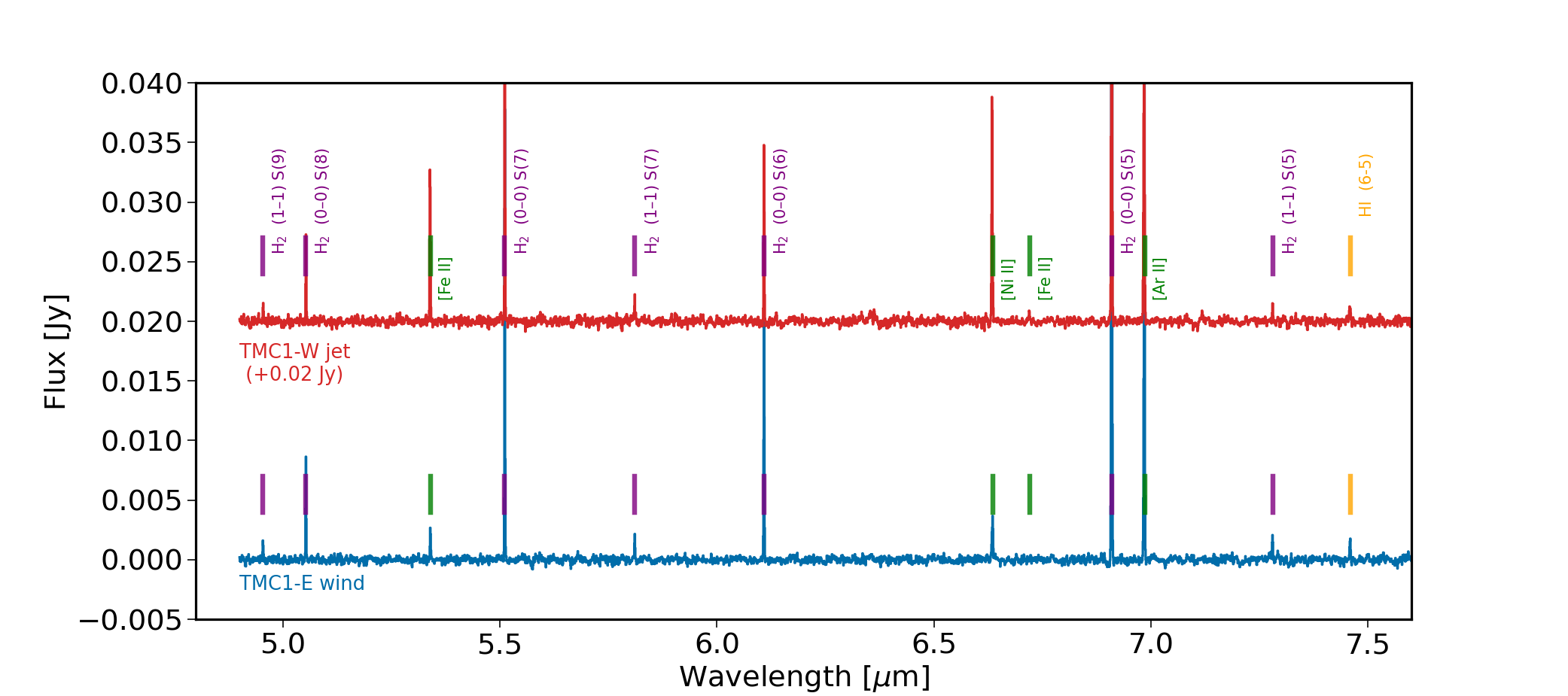}
        \includegraphics[width=0.95\textwidth]{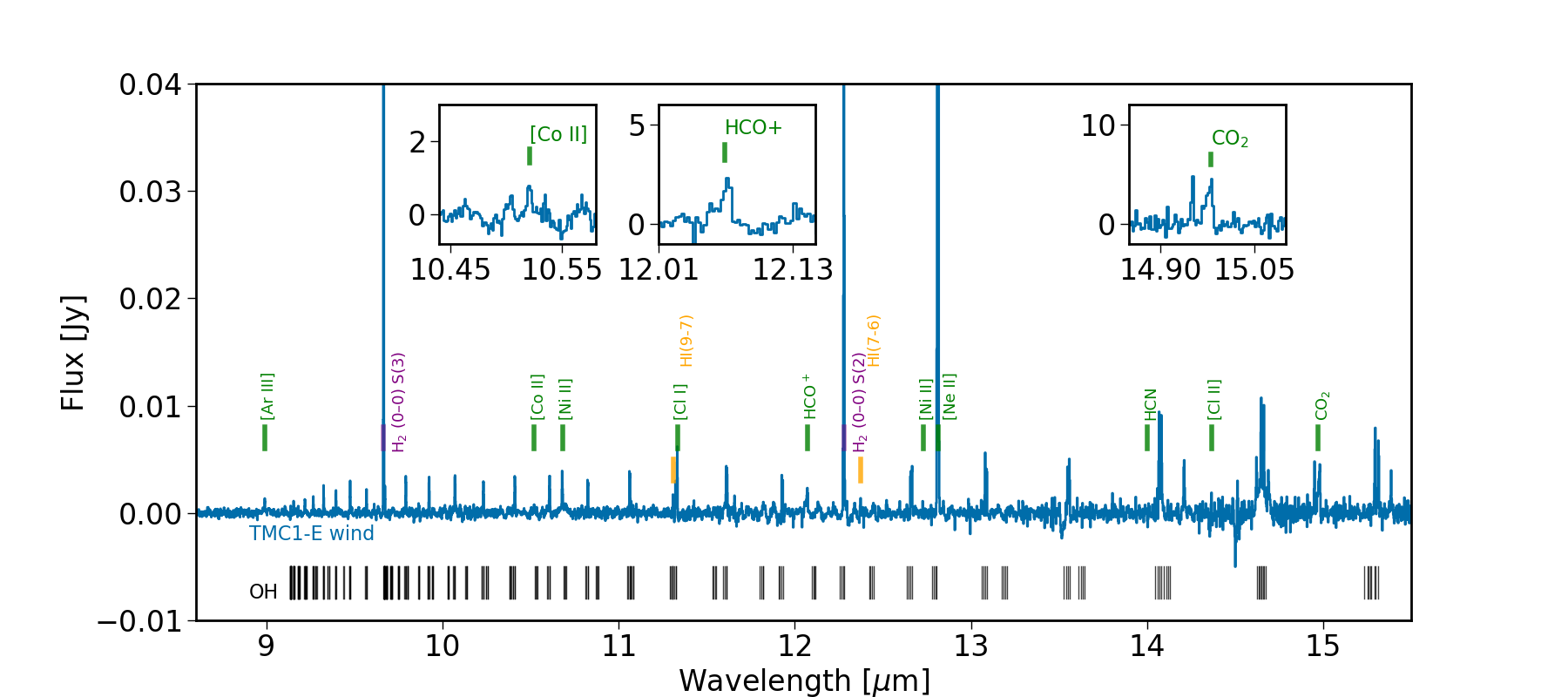}

    \caption{MIRI-MRS spectra from the outflow positions. Spectra were continuum subtracted using a spline fit to line-free regions.
    The H$_2$ transitions are shown in purple, the hydrogen recombination line in yellow, and fine structure atomic lines in green. {\it Top:} TMC1-W jet (red) and TMC1-E wind (blue).  {\it Bottom:} TMC1-E wind spectra. OH catalog emission line positions are shown as black markers \citep{Yousefi.Bernath.ea2018,Tabone.vanHemert.ea2021}. Insets: [Co II] 10.521 $\mu$m, HCO$^+$ 12.069 $\mu$m, and CO$_2$ 14.98 $\mu$m lines identified in the spectra. Flux in the insets converted to mJy.}
    \label{fig:spectra_overview_outflows}
\end{center}
\end{figure*}

\section{Observations}

\subsection{JWST-MIRI}
JWST MIRI-MRS observations were obtained on March 4, 2023, as part of the JWST Observations of Young ProtoStars (JOYS) Guaranteed Time Observations (GTO) PID 1290 (PI: E. F. van Dishoeck). Three sub-bands were integrated with 36 groups in the FASTR1 integration mode with two dithers in a negative direction optimized for the extended source. The integration time per sub-band was 200s, so 600s to obtain full spectrum from 4.9 $\mu$m to 28.6 $\mu$m with a spectral resolving power of R~1500-3500. The target position was 4$^h$41$^m$12$^s$.7 +25${^\circ}$46\arcmin34\farcs8 (J2000). Integration on the source was preceded by the off-source background observation centered on coordinates 4$^h$39$^m$32.$^s$3 +26${^\circ}$05\arcmin14\farcs5 (J2000). The background observation was obtained with 72 groups in the FASTR1 integration mode, with no dithering, resulting in the same total integration time as the science target.

The data were reduced with the JWST pipeline v.1.11.2 \citep{Bushouse.Eisenhamer.ea2023} using the calibration reference data system (CRDS) context file \texttt{jwst1100.pmap}. The raw telescope data were processed with the \texttt{Detector 1} pipeline with default settings. Detector calibrated files are constructed in the \texttt{Spec2} step, including fringe flat for extended sources (Mueller et al. in prep.) and residual fringe correction (Kavanagh et al. in prep.). In this step, the background image is subtracted from the detector image using dedicated background rate files. At this step, a bad-pixel routine is applied to calibrated files using the Vortex Image Processing (VIP) package \citep{Christiaens.Gonzalez.ea2023}. Final data cubes for each channel and sub-band are created at \texttt{Spec3} pipeline step, with the outlier rejection step turned off since this is handled by a custom-made bad-pixel routine. The absolute flux calibration uncertainty of MIRI-MRS is estimated from flight performance at 5.6 $\pm$ 0.7 $\%$ \citep{Argyriou.Glasse.ea2023}.

\subsection{ALMA}

For the analysis, we also used archival ALMA data. CO $J$$=$2--1 230.538 GHz data and SO  $J$$=$$5_6$--$4_5$ 219.9494 GHz with a synthesized beam (resolution element) of 0\farcs41 $\times$ 0\farcs29 are from project 2017.1.01350.S (PI: Tychoniec). CO 2--1 and SO $5_6$--$4_5$ has 0.63 km s$^{-1}$ and 0.33 km s$^{-1}$ velocity resolution data, respectively. CN  $J$$=$2--1 225.632 GHz data with a synthesized beam of 0\farcs57 $\times$ 0\farcs38 and velocity resolution of 0.08 km s$^{-1}$ are from 2017.1.01413.S (PI: van 't Hoff).

Due to the higher sensitivity of the latter dataset, we used it to construct a continuum image at 1.3 mm from line-free spectral regions for further analysis. We used pipeline-calibrated data images obtained from the ALMA archive for both projects.

\section{Results}

The MIRI-MRS cubes of the TMC1 protostellar system reveal two continuum point sources. Across the MIR spectrum, we also detect several spatially extended features associated with emission lines. An overview of the system is presented in Fig. \ref{fig:representative_regs}. 

\subsection{MIRI-MRS Spectra}
\label{sec:spectra}
This section analyzes spectral information extracted from positions indicated in Fig. \ref{fig:representative_regs}. The extracted regions were selected to cover both continuum peaks associated with two protostars of the TMC1 system and the representative outflow positions. These positions allow for optimal disentanglement of the outflow components. The extraction aperture at the point source was set to $r_{a} = 0.033 \times \lambda + 0\farcs106$, which at 5 $\mu$m is 0\farcs271 and at 20 $\mu$m is 0\farcs766. This is twice the size of the FWHM of MIRI-MRS PSF as characterized by \cite{Law.Morrison.ea2023}. At the outflow positions with prevalent extended emission, we used a circular aperture of 0\farcs25 radius to ensure that equal regions are extracted at different wavelengths.

The location of extracted spectra for continuum peaks and outflow positions are indicated by star symbol and white circles in Fig.\ref{fig:representative_regs}, respectively.
After spectral extraction, additional fringe correction is applied to the spectra \citep[][Kavanagh et al. in prep.]{Gasman.Argyriou.ea2023}. A Gaussian profile was fitted to each identified emission line in the spectra and the flux value of the line was estimated by integrating the region under the Gaussian. The line intensity was then obtained by dividing the final flux value with the extraction aperture area.

To obtain the true line intensity, an extinction correction should be applied, accounting for attenuation of the flux by the envelope in which the forming star is embedded. For TMC1 protostar, a visual extinction of $A_V$~=~49.2 mag is reported in \cite{Connelley.Greene2010} based on spectral energy distribution shape in the 0.8 to 2.4 $\mu$m range. \cite{Fiorellino.Tychoniec.ea2023} found a value of 36.77–35.70 mag for birth line and at 1 Myr, respectively. However, both those methods assume that the flux comes from a single star; therefore, these estimates are unreliable. In \cite{Apai.Toth.ea2005} near-IR colors of two components are evaluated separately, and TMC1-W and TMC1-E are found to have $A_V$ of 43 and 37 mag, respectively. However, the colors of protostellar sources can be significantly affected by the presence of the disks that both protostars have, as evident in submillimeter continuum images \citep{Harsono.Joergensen.ea2014, vantHoff.Tobin.ea2018}, and the models of color correction developed for T Tauri stars can underestimate the extinction when used in the case of younger protostars \citep{Doppmann.Greene.ea2005}.

Given the absence of reliable extinction estimation for this source, we use MIRI-MRS data to estimate $A_V$~ with different methods for on-source and off-source positions. On protostars, we measure silicate feature depth after the continuum subtraction using a third-order polynomial function. This process is presented in the Appendix \ref{sec: extinction}. We find $\tau_{\textrm 9.7}=1.30$ and $\tau_{\textrm 9.7}=1.95$ for TMC1-W and TMC1-E, respectively. 

Conversion of $\tau_{\textrm 9.7}$ to $A_V$ bears large uncertainties, with factors of A$_V=18.5$ $\times$ $\tau_{\textrm 9.7}$ used for ISM and A$_V$ = 9 $\times$ $\tau_{\textrm 9.7}$ used for dense clouds \citep{McClure2009, Weingartner.Draine.ea2001}. This is due to the total-to-selective extinction ($R_V$) in the two environments \citep{Chiar.Ennico.ea2007}.  
\cite{McClure2009} shows that extinction values toward star-forming regions are in between those provided for the galactic center and diffuse ISM. Therefore, we use those two limits as a range for the extinction estimation. This results in $A_V$=$11.7$--$24.0$ mag and $A_V$=$17.6$--$36.1$ mag for TMC1-W and TMC1-E, respectively.

%With A$_V$/A$_K$ = 7.75 for R$_V$ = 5 following the method described in \cite{McClure2009}, we get A$_K$ = 3.10 and 4.65 mag for TMC1-W and TMC1-E, respectively. The relation between A$_\lambda$ and A$_K$ is provided following the extinction curve in \cite{McClure2009}. The important caveat is that we do not have reliable extinction measurements across the envelope. Therefore, there is some uncertainty on how variable the extinction is at different positions.

At the off-source position, where no continuum is present to measure the depth of the silicate feature, we use H$_2$ emission lines. We fit simultaneously for the rotational temperature and extinction value \citep{Barsony.WolfChase.ea2010, Federman.Megeath.ea2023}. Details are provided in the Appendix \ref{sec: extinction}. We only use v=0--0 transitions because other non-thermal effects can contribute to the ratio of fluxes between v=0--0 and v=1--1 transitions \citep{Timmermann.Bertoldi.ea1996}. The value measured at TMC1-E blueshifted wind position $A_V=20.8 \pm 3.1$ mag and at TMC1-W blueshifted jet $A_V=23.1 \pm 2.8$ mag. We use these values to correct for line fluxes measured at both positions. 

We note that considering the ranges of the extinction values found for the on-source and outflow positions, there is a possibility that the extinction values in the outflow would be higher than on-source. This is unexpected as much larger extinctions are typically measured close to the protostar given strongly decreasing gas density with increasing envelope radius \citep{Jorgensen.Schoier.ea2002}. However, it should be considered that dust grain properties change across the envelope, with rapid dust size increase close to the protostar \citep{Cacciapuoti.Macias.ea2023}, making the two extinction values challenging to compare as the total-to-selective extinction depends strongly on dust size \citep{Chiar.Ennico.ea2007}.

%The values are consistent with the ranges found for the silicate absorption features on the protostellar continuum.

\subsubsection{Spectra at source positions}

Figure \ref{fig:spectra_overview} presents spectra extracted from the central position of the protostars on the MIRI-MRS detector. From Channel 3 short onward ($>$ 11.55 $\mu$m), the regions of spectral extractions significantly overlap; however, at shorter wavelengths, it is possible to disentangle the emission from the two sources and discuss their differences. Both sources have spectra characteristic of embedded protostars, with a spectral energy distribution (SED) rising with wavelength. The spectra show strong emission lines of molecular H$_2$, atomic and ionized components, along with deep silicate and ice absorption features, such as H$_2$O (6 $\mu$m), CH$_3$OH (7 $\mu$m), silicates  (9 $\mu$m), and CO$_2$ (15 $\mu$m) \citep{Ehrenfreund.Boogert.ea1997, Keane.Tielens.ea2001, Zasowski.Kemper.ea2009, Bottinelli.Boogert.ea2010, Yang.Green.ea2022, Rocha.Rachid.ea2022}.

The spectra also reveal weaker emission lines towards both sources, such as emission from CO molecular gas in the 5 $\mu$m spectral region and H$_2$O emission lines detected in 5--8 $\mu$m range (see bottom-right inset in Fig \ref{fig:spectra_overview}). In the same inset, we compare spectra towards both targets with CO and H$_2$O slab gas spectra at 1000 K under the assumption of LTE. In the 9--20 $\mu$m spectral range, OH emission is prominently detected towards TMC1-E, with only a hint towards TMC1-W (which could also be the result of emission from TMC1-E included in the aperture). CO$_2$ gas-phase emission lines at 14.98 $\mu$m are detected towards both sources on top of the deep ice absorption feature. 

Molecular H$_2$, atomic, and ionized emission lines are detected towards both targets. However, verifying whether the emission comes from the inner disk, protostar, or outflow is difficult. The continuum-subtracted maps of the H$_2$ and [Fe II] line in Fig. \ref{fig:representative_regs} clearly indicate that these lines originate mainly off-source. Therefore, while the contribution of the central region cannot be discarded, we assume that most of H$_2$ and ionized emissions are coming from the winds or jets of both protostars. 

The exceptions are atomic hydrogen recombination lines, which are brightest at the on-source apertures; see top-left inset in Fig \ref{fig:spectra_overview}. In Table \ref{tab:hydrogen_both}, we report extinction-corrected line intensities of the atomic hydrogen transitions detected towards the two binary components of the TMC1 system. We identify six transitions of atomic hydrogen toward both protostars, with line intensities significantly higher towards TMC1-W. These lines are expected to trace high temperature and density environments and are usually associated with stellar accretion  \citep{Rigliaco.Pascucci.ea2015,Alcala.Manara.ea2017}. However, not all MIR lines have derived relations between the accretion rate and line fluxes. The best one to date is the HI 7--6 line for which the relation is provided in \cite{Rigliaco.Pascucci.ea2015}.

In Section \ref{sec:spectra}, we provided ranges for the extinction values for TMC1-W and TMC1-E based on the different correlations found for different grain properties. It is safe to assume that the dust grains should have similar properties toward both protostars. Therefore, the relative error on extinction should not be as large as the range of values provided. In the subsequent analysis, we assume the mean value from both ranges. This is $A_V=$27 mag and $A_V=$18 mag for TMC1-E and TMC1-W, respectively.

We obtain the extinction corrected line luminosity for this line to be $3.8\times 10^{-7}\ \lsun$ and $1.8\times 10^{-7}\ \lsun$,  for TMC1-W and TMC1-E, respectively. Applying the relation between HI 7--6 lines and accretion luminosity derived from H$\alpha$ provided in Eq. 1 in \cite{Rigliaco.Pascucci.ea2015},  and accounting for the uncertainty on this relation, this corresponds to the accretion luminosity of 2.3$^{+9.1}_{-1.8}$ $\times$ 10$^{-4}$
and $0.5^{+3.5}_{-0.48}$ $\times$ 10$^{-4}$ $\lsun$, respectively. Next, using Equation 11.5 from \cite{Stahler.Palla2005} (see also \citealt{Gullbring.Hartmann.ea1998}), we get:

\begin{equation}
\label{eq1}
    \dot{M}= L_{\rm acc}GM_{*}R_{*}^{-1},
\end{equation}

% \begin{equation}
%\label{eq1}
%\resizebox{0.3\linewidth}{!}{^\dot{M} =  L_{\rm acc}GM_{*}R_{*}^{-1},   } 
%\end{equation}

and taking the range of stellar properties of TMC1, namely, M$_\star$=0.15-0.20 $\msun$ and R$_\star$=3.0-2.48 $R_\odot$ for birthline and 1 Myr, respectively, from self-consistent modeling in \cite{Fiorellino.Tychoniec.ea2023}, we find the accretion rate of TMC1-E to be $(0.02-0.43) \times 10^{-11}\ \msun\ {\rm yr}^{-1}$ and of TMC1-W to be $(0.16-14.8) \times 10^{-11}\ \msun\ {\rm yr}^{-1}$. This range includes uncertainties on the accretion luminosity.

There are several caveats to this analysis. First, the atomic hydrogen emission lines do not have to trace exclusively accretion but can also be detected in the protostellar jet and wind, which is evident especially in recent spatially resolved JWST observations of protostars \citep{Federman.Megeath.ea2023, Harsono.Bjerkeli.ea2023}. While we do not detect significant hydrogen emission lines in the jet, they could still contribute to the emission close to the source.

Second, stellar properties in \cite{Fiorellino.Tychoniec.ea2023} are obtained under the assumption of a single star dominating the flux and, therefore, more accurate for the brighter TMC1-W, with TMC1-E properties remaining uncertain. Additionally, the flux obtained for the HI (7-6) line towards TMC1-E might be contaminated by contribution from TMC1-W, given the large PSF at 12.37 $\mu$m (FWHM $\sim$0\farcs5). Given that extraction region size at this wavelength is 0\farcs5, and the distance between the sources is 0\farcs65, there can be significant contribution from TMC1-W. Comparing ratios of HI lines between the sources, we find that the HI (9-6) line, at 7.5 $\mu$m,  is 17 times brighter for TMC1-W compared with TMC1-E, while the HI (7-6), at 12.37 $\mu$m is only 2 times brighter. This further suggests increasing the contribution of TMC1-W flux to TMC1-E aperture with wavelength. The considerable uncertainty on the extinction correction also adds to the difficulty when comparing the two sources. In summary, while highly uncertain, these measurements indicate the stellar accretion conditions for both targets, tentatively pointing to TMC1-W being a stronger accretor in the system. 

\begin{figure}
    \centering
    \includegraphics[width=0.49\textwidth]{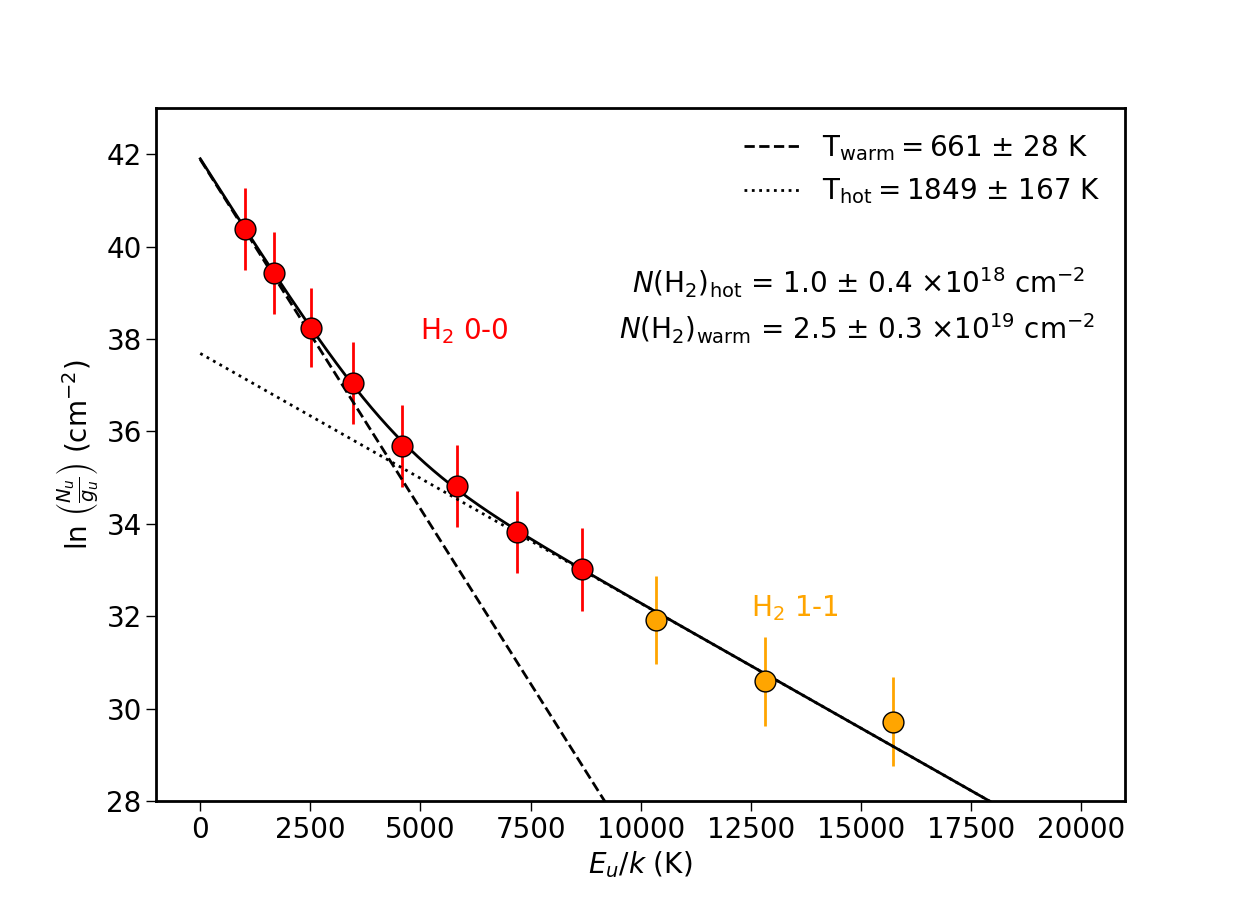}
    \caption{H$_2$ rotational diagram obtained with line intensities measured at the TMC1-E wind position. All detected H$_2$ transitions are presented, but only v=0-0 (red) transitions are used to fit the temperature. }
    \label{fig:h2rot}
\end{figure}

\subsubsection{Spectra at the outflow positions}

In Fig. \ref{fig:spectra_overview_outflows} (top panel), we show 5--7.6 $\mu$m spectra extracted from outflow positions for TMC1-E and TMC1-W outflows. At longer wavelengths, the spectra are affected by contribution from the central source PSF. Hence, we selected positions with a significant distance from the protostar for spectral extraction but close enough to be within the MRS footprint in all channels. The spectra plotted in Fig. \ref{fig:spectra_overview_outflows} were extracted with an aperture size of 4 $\times$ PSF FWHM to minimize the effect of undersampling of the spectral information in the MRS detector, while for the flux measurements,  we used a fixed circular aperture size of 0\farcs25 radius.

 H$_2$  pure rotational transitions are prominently detected in our data, primarily towards the TMC1-E wind position. The detection of  H$_2$  from TMC1-W is likely the result of contamination with the western side of the TMC1-E wind (see Fig. \ref{fig:representative_regs}). The line intensities of H$_2$ lines towards TMC1-E outflow are reported in Table \ref{tab:h2_tmc1e}. 

To analyze the physical conditions associated with H$_2$ emission, we construct a rotational diagram using \texttt{pdrtpy} package \citep{Pound.Wolfire2008}. Figure \ref{fig:h2rot} shows a rotational diagram produced using the measured fluxes corrected for the visual extinction (see Section \ref{sec:spectra}). We only used v=0-0 transitions to quantify the rotational temperature. By fitting simultaneously two linear functions to the data, the negative inverse of the slope provides the rotational temperature of two separate components: a warm component at 661 $\pm$ 28 K and a hot component at 1849 $\pm$ 167 K. Taking the intercept of the linear functions multiplied by the partition function \citep{Herbst.Beckwith.ea1996}, we find the total column density of the two components: 2.5 $\pm$ 0.3 $\times$ 10$^{19}$ cm$^{-2}$ for the warm and 1.0 $\pm$ 0.4 $\times$ 10$^{18}$ cm$^{-2}$ for the hot component, with hot gas contributing only 1$\%$ to the total column density. We do not provide a similar analysis for the TMC1-W jet region since any detected H$_2$ emission at this position is likely coming from the TMC1-E outflow.

While the molecular CO and H$_2$O have strong emissions on the source positions, they are not detected in the outflow. Other molecules, namely, HCO$^{+}$, OH, and CO$_2$, appear to be emitted from the outflow (see bottom panel of Fig. \ref{fig:spectra_overview_outflows}); however, the contamination from the source position cannot be excluded. The lines are not bright enough to map them with a sufficiently high signal-to-noise ratio (S/N). We detected several atomic and ionized emission lines at the outflow positions, namely,  [Fe II], [Ne II], [Ne III], [Ni II], [Ni I], [Ar III], [Ar II], [S I], [Cl II], [Cl I]. We also tentatively detected [Co II] (Fig. \ref{fig:spectra_overview_outflows}). We found only one instance of detection of this species in the protostellar jet in \cite{Giannini.Antoniucci.ea2015}. Due to the lack of other detected transitions, this identification is tentative. However, this is a relatively low-energy transition to the ground level with \eup\ of 1368 K and a high $A_{ij}$ coefficient of $2.2 \times 10^{-2}$ s$^{-1}$ \citep{Nussbaumer.Storey1988}, so this is a plausible association that requires more investigation. Table \ref{tab:ions_both} reports the extinction-corrected line intensities for both outflows. Table \ref{tab:line_rations} reports the relevant line ratios for both outflow positions.

\begin{figure*}
    \centering
    \includegraphics[width=0.92\textwidth]{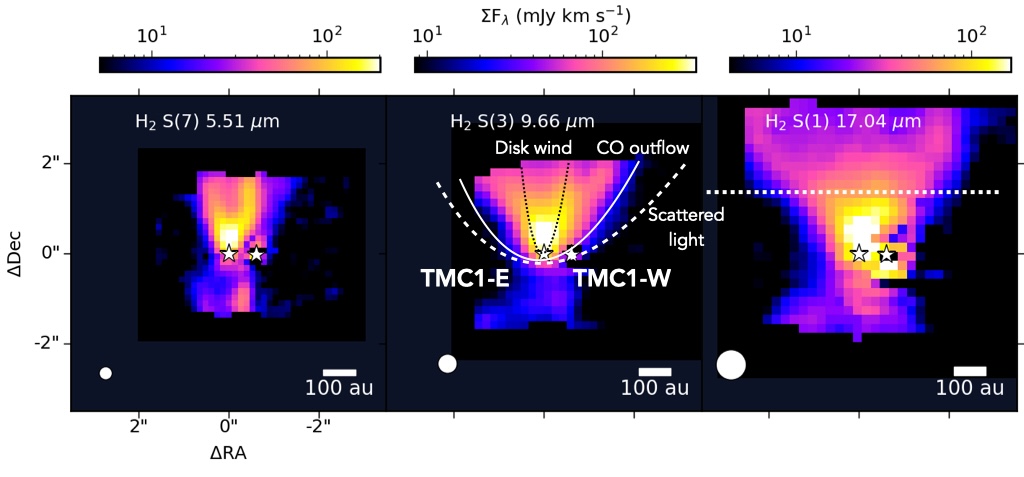}
    \caption{Three pure rotational transitions of H$_2$ from left to right: S(7) at 5.51$\mu$m, S(3) at 9.66$\mu$m, and S(1) at 17.04$\mu$m. Integrated emission maps are created from the cubes after subtracting the local continuum. The positions of the protostars measured from the line-free region at 5$\mu$m are shown with white stars. Maps are centered on the position of the TMC1-W source. The horizontal dotted line in the right panel indicates the cut for which different H$_2$ 0-0 S bands are shown in Fig. \ref{fig:cutout}. The physical components of the system are illustrated in the middle panel. The disk wind is indicated based on H$_2$ S(7), CO outflow is based on the ALMA observations, and scattered light on the HST 1.6 $\mu$m image\ in Fig. \ref{fig:ALMA_maps}. Bottom-left corners: MIRI-MRS empirical FWHM of PSF \citep{Law.Morrison.ea2023} indicated as a white circle.}
    \label{fig:h2_integrated_maps}
\end{figure*}

\subsection{Emission maps:\ MIR characterization of the TMC1 system}
This section presents the emission maps of lines detected toward the TMC1 binary system.
Figure \ref{fig:h2_integrated_maps} shows H$_2$ 0-0 S(1), S(3), and S(7) lines, while Fig. \ref{fig:ion_integrated_maps} shows [Fe II], [Ni II], [Ar II], [Ne II], and [Ne III]. To produce the continuum-free emission line maps, we subtracted the line-free part of the spectral cube close to the line from the integrated emission map centered on the line wavelength. We only consider lines with a sufficiently high S/N to separate the extended emission from the continuum and the noise. Therefore, not all the detected lines are present on the maps. The spectral resolution of the MIRI-MRS \citep[R$\sim1500-3500$][]{Labiano.Argyriou.ea2021} allows for the production of channel maps of the emission lines for the high-velocity gas in protostellar jets. The maps are presented in \ref{fig:chanmaps_v1}.

\subsubsection{Molecular hydrogen emission lines} 

Molecular hydrogen (H$_2$) is an excellent tracer of the bulk of the outflowing gas from protostars; however, due to its lack of permanent electric dipole moment, it can only be detected through $\Delta J=2$ rotational transitions with relatively high excitation energies (\eup $>$ 500 K), thereby tracing the heated gas \citep{Neufeld.Nisini.ea2009}. We see that  H$_2$ emission originates from TMC1-E, with the apex of the parabolic shape of the emission pointing to this source. Figure \ref{fig:h2_integrated_maps} shows the rotational S-branch transitions of $v=0-0$ state of H$_2$. While we cannot completely exclude the H$_2$ coming also from the TMC1-W (see fainter broad component in H$_2$ S(7) line in Fig. \ref{fig:h2_integrated_maps}), we do not see any further extent of this wind at longer wavelengths, especially at S(1), where the wide-angle wind is completely dominated by the TMC1-E wind. Therefore, a significant contribution to H$_2$ emission from TMC1-W is unlikely.

  H$_2$ emission is present inside the cavity and shows progressively larger opening angles with decreasing upper energy levels. Additionally, two components can be observed: a narrow V-shaped emission and a much broader parabolic shape (annotated on H$_2$ S(3) map). We interpret those as tracing the less collimated wind and the shocks filling the whole outflow cavity wall, respectively. It can be seen that the emission from the wind component remains narrow and compact across all H$_2$ transitions, while the outflow cavity component becomes more prominent with decreasing \eup. Maps reveal outflow emission on the south side of the source as well. However, this emission is much fainter, likely because this is the redshifted side of the flow, hidden on the other side of the protostellar envelope of the TMC1 system.

\begin{figure}
    \centering
    \includegraphics[width=0.4\textwidth]{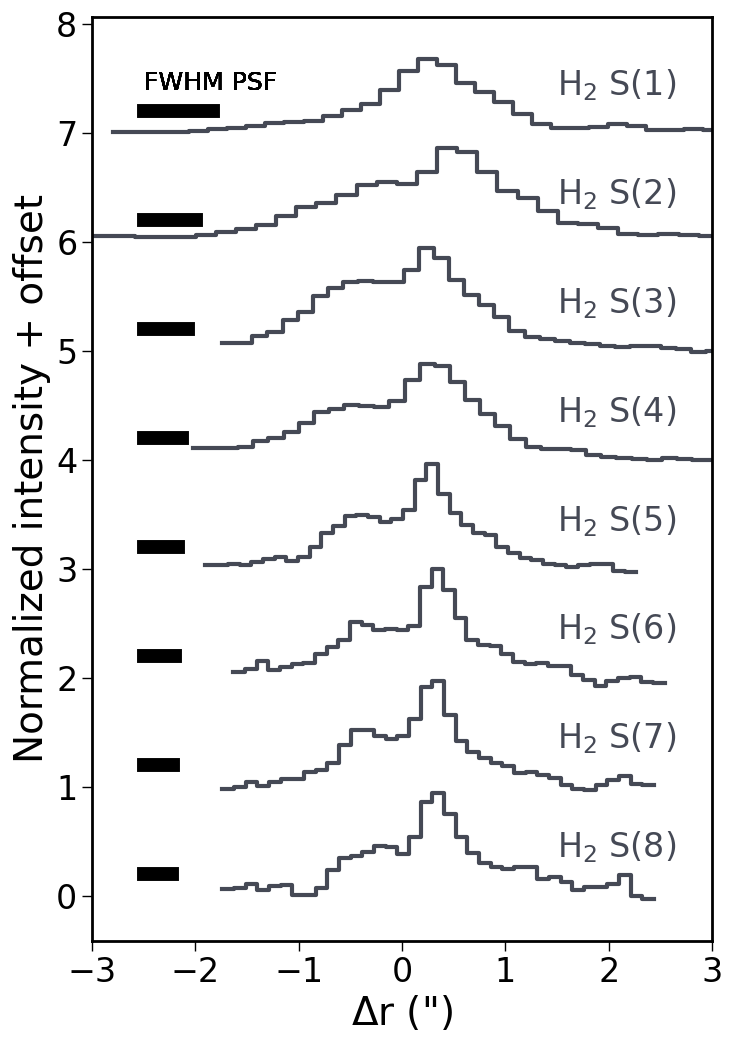}
    \caption{Radial cuts across the TMC1-E outflow at 1\farcs5 distance from the protostar for each of the H$_2$ (0--0) transitions, with upper energy levels increasing from top to bottom. The reference is set to the protostar position. The position of the cut is shown in Fig. \ref{fig:h2_integrated_maps}. The black bar shows the size of the FWHM of instrumental PSF, as described in \cite{Argyriou.Glasse.ea2023}.}
    \label{fig:cutout}
\end{figure}

To illustrate the increasing opening angle of the H$_2$ emission, in Fig. \ref{fig:cutout}, we show radial cuts across the blueshifted outflow of TMC1-E. For the high \eup\ lines, there is a clear two-peaked feature that can be associated with the disk wind, and the whole extent of the wind does not exceed 1\arcsec, while at lower \eup\ broader component is also detectable extending up to 2\farcs5.

\subsubsection{Ionized and atomic species}

Atomic and ionized species in MIR can trace various physical components of the systems. Spatially unresolved {\it Spitzer}-IRS observations identify the origin of lines in jets, disks, and photodissociation-like environments in the outflow cavity walls \citep{Hollenbach.Gorti2009, Lahuis.vanDishoeck.ea2010}. The MIRI spatial resolution is invaluable in shedding new light on the origin of those lines. 

\begin{figure*}
    \centering
    \includegraphics[width=0.92\textwidth]{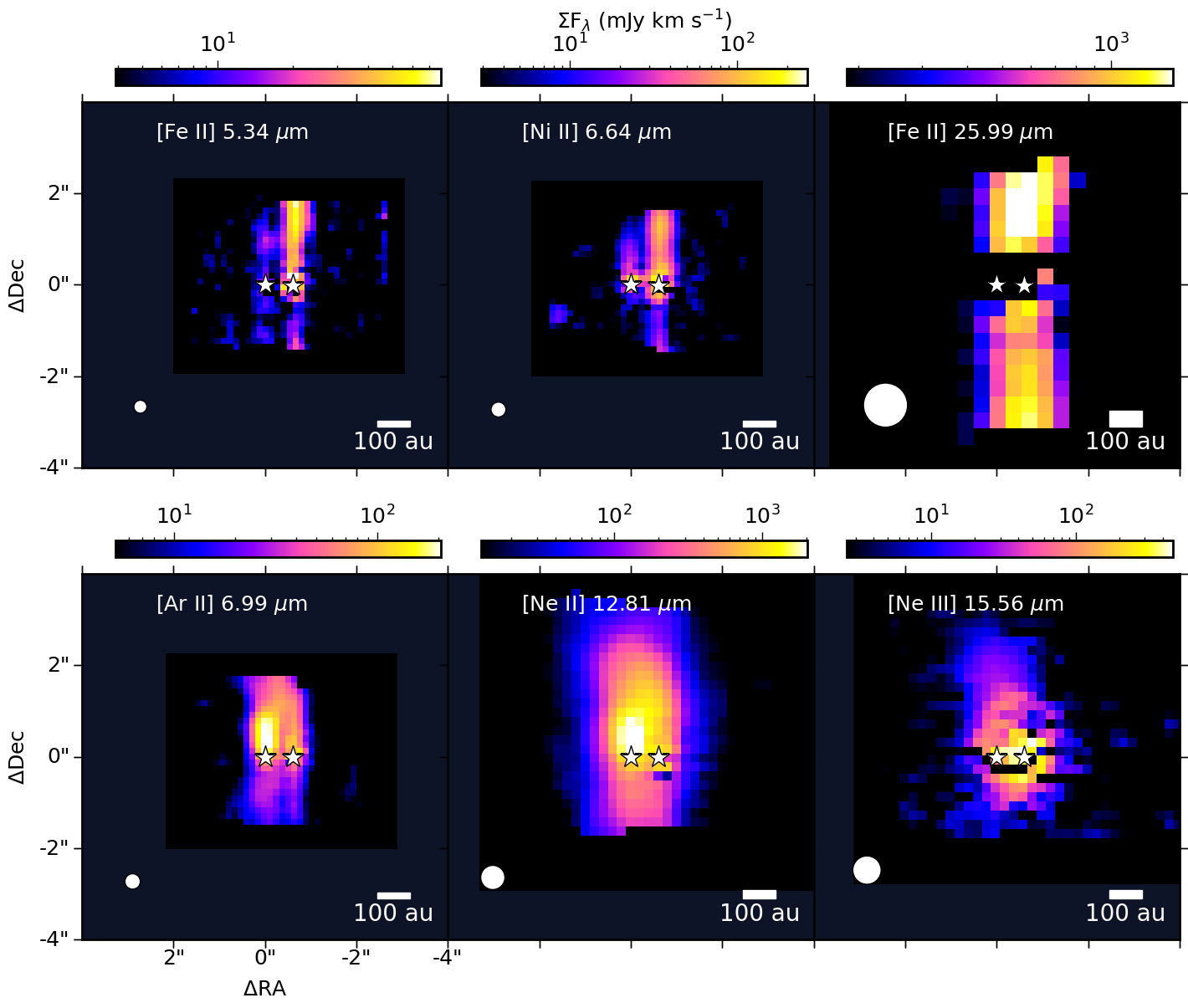}
    \caption{Integrated emission maps of selected emission lines of ionized species. Integrated emission maps are created from the cubes after subtracting the local continuum. Positions of the protostars measured from the line-free region at 5$\mu$m are shown with white stars. Maps are centered on the position of the TMC1-W source. Bottom-left corners: MIRI-MRS empirical FWHM of PSF \citep{Law.Morrison.ea2023} indicated as a white circle.}
    \label{fig:ion_integrated_maps}
\end{figure*}

In Fig. \ref{fig:ion_integrated_maps}, we present key atomic and ionized tracers detected toward the TMC1 system. Based on their appearance, they can be divided into two categories: refractory and noble gas ions. The refractory ions [Fe II] and [Ni II] show a collimated jet (radially unresolved) appearance. They are much brighter in the jet from TMC1-W where H$_2$ lines are undetected, while only weakly detected towards source TMC1-E, where the H$_2$ emission is prominent. On the other hand, noble gas ions [Ar II], [Ne II], and [Ne III] are present in both sources, but they are distinctly brighter in the TMC1-E outflow. They also appear moderately extended in the radial spatial direction, contrasting the collimated refractory ions. This is particularly clear in the channel maps where [Ni II] and [Ar II] are compared (Fig. \ref{fig:chanmaps_v1}). 

Channel maps reveal velocity information contained in the data obtained with MIRI. In the case of the TMC1-W jet both [Ni II] and [Ar II] are detected at velocities up to 200 km s$^{-1}$, while gas at lower velocities is detected in the case of TMC1-E, especially the radially extended component of the TMC1-E wind is not detected at velocities exceeding 100 km s$^{-1}$.

\begin{figure*}
    \centering
    \includegraphics[width=0.96\textwidth]{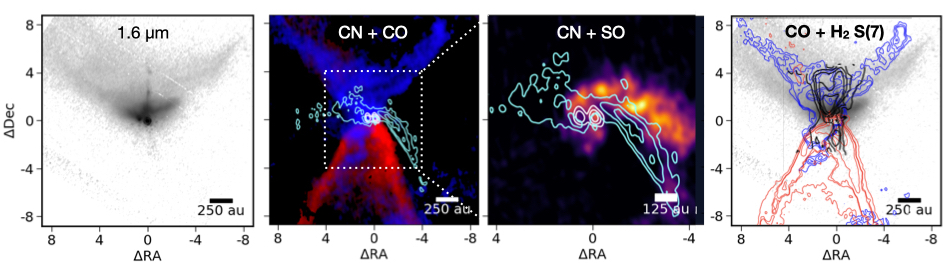}
    \caption{{\it } {\it HST} NICMOS 1.6 $\mu$m image of the TMC1 system (far-left).  ALMA CO (2-1) blueshifted and redshifted integrated emission in colorscale, ALMA 1.3 mm continuum in white contours, and CN peak emission map (moment 8)  in cyan contours (center-left). {\it } ALMA SO (6-5) integrated emission map in colorscale (center-right).  NICMOS 1.6 $\mu$m image in colorscale with ALMA CO(2-1) overlaid in red and blue contours, and H$_2$ S(7) moment 0 map in black contours (far-right).}
    \label{fig:ALMA_maps}
\end{figure*}

%\subsection{Atomic hydrogen emission}

%Because it is difficult to separate the emission from two components, we will assume that the TMC1-W is the dominant source of the flux for those lines, as is evident by examining the extraction apertures: the maximum flux is achieved when extracting the spectrum from the position of the TMC1-W protostar. 

\subsection{Molecular emission at submillimeter wavelengths and scatter light near-IR emission}

For young stellar objects (YSOs), putting the MIR observations in the context of colder gas tracers detected in the (sub-)millimeter range is particularly insightful. This is because those objects are still embedded in their natal envelope and many are still accreting material from the cloud onto the disk. Additionally, colder molecular gas will be released in entrained outflow. In particular, the low-$J$ rotational transitions of CO are excellent tracers of the low-velocity outflows \citep[e.g.,][]{Arce.Borkin.ea2010,Plunkett.Arce.ea2013,vanderMarel.vanDishoeck.ea2013}. The (sub-)millimeter range is also less affected by extinction. The velocity resolution of ALMA below 1 km s$^{-1}$  could provide valuable insights into interpreting velocity unresolved information from MIRI-MRS.

Near-IR or optical images of embedded systems at shorter wavelengths show scattered light coming from the protostellar envelopes or outflow cavity walls. These observations are helpful as they inform about the geometry of the system and the presence of small grains.

We show an archival {\it Hubble} Space Telescope NICMOS 1.6 $\mu$m image of the TMC1 system in Fig. \ref{fig:ALMA_maps} (left panel), presented in \cite{Apai.Toth.ea2005} and \cite{Terebey.VanBuren.ea2006}. The extended emission on the northern side of the protostar is attributed to the scattered light at the outflow cavity walls. The scattered light presents a parabolic shape with an apex at the TMC1-W protostar. A collimated jet is also emerging from the TMC1-W source, with [Fe II] line at 1.64 $\mu$m, likely to be dominating the jet emission in the F160W imaging filter \citep{Terebey.VanBuren.ea2006}. Scattered light most likely shows the boundaries of the outflow cavity walls. Since they appear broader than the CO and H$_2$ outflow currently observed towards TMC1-E, this suggests that in the past, both TMC1-W and TMC1-E contributed to shaping the outflow cavity walls; however, as currently observed, there is no visible wide-angle wind from TMC1-W.

We also note that no scattered light is detected with MIRI, in contrast with Class 0 sources also observed at MIR \citep{Federman.Megeath.ea2023}. This highlights the more evolved nature of the TMC1 system compared with Class 0 protostars.

In Fig. \ref{fig:ALMA_maps}, we present ALMA maps of CO 2--1 (central panel), SO $5_6$--$4_5$ (right panel), and CN 2--1 (contours in the central and right panel) from \cite{Tychoniec.vanDishoeck.ea2021}. CO emission traces the entrained outflow material piling up at the outflow cavity walls of TMC1-E. Notably, CO emission has a narrower opening angle compared with the scattered light image, and the center of the symmetry of CO outflow appears to be aligned with the TMC1-E. \cite{Terebey.VanBuren.ea2006} found the best fit for the system's inclination of 50$\degree$ based on the scattered light modeling and already identified  the opening angle of the scattered light as being larger than CO outflow.

The CN emission is narrow and not co-spatial with the outflow cavity walls, as evident from the comparison with the CO map. The velocity gradient is consistent with infall motion and inconsistent with outflow, and velocity dispersion is small $<$0.2 km s$^{-1}$ (Fig. \ref{fig:cnmom1}). Also, SO highlights the gas likely associated with weak shocks, which can inform where the infalling material interacts with the envelope and disk \citep{Sakai.Sakai.ea2014}. It may also be tracing the young, embedded disk not associated with accretion, which can explain its presence towards both TMC1-W and TMC1-E \citep{Harsono.Joergensen.ea2014}. It is worth noting that even though the SiO 5--4 transition is covered in the ALMA data, it was not detected towards TMC1. This further highlights the more evolved nature of the protostars. High-velocity SiO jets are commonly observed towards the youngest, Class 0 protostars \citep{Hirano.Ho.ea2010,Tychoniec.Hull.ea2019, Podio.Tabone.ea2021}, where molecules can form efficiently in the jets due to high densities \citep{Tabone.Godard.ea2020}. As the outflow evolves, the mass loss rate decreases, velocities increase, and the jet becomes progressively more ionized \citep{Nisini.Santangelo.ea2015}.

All observed features are consistent with an accretion streamer feeding the system \citep{LeGouellec.Hull.ea2019, Pineda.SeguraCox.ea2020, ValdiviaMena.Pineda.ea2022}. We show CN data as an example of the accretion streamer because (for those lines) the highest spectral resolution data are available. However, closer to the source, there is significant contamination from the envelope and disk material. Moreover, SO emission aids in pinpointing currently shocked gas, from which we can see that the streamer appears infalling mostly onto the disk around the TMC1-E protostar.

%In Fig. \ref{fig:ALMA_maps} integrated intensity (moment 0), intensity weighted velocity (moment 1), and velocity dispersion (moment 2) maps of CN (transition) are presented. Emission is narrow and not co-spatial with the outflow cavity walls, velocity gradient consistent with infall motion and inconsistent with outflow, and velocity dispersion is small $<$0.2 km s$^{-1}$.  All observed features there are consistent with an accretion streamer \citep{Pineda.SeguraCox.ea2020}. 

%For comparison, we show the scattered light emission 1.6 $\mu m$ and CO (2-1) emission map (Fig. \ref{fig:nicmos_alma}). Those tracers are delineating the outflow cavity walls.

\section{Discussion}

MIRI-MRS observations have revealed for the first time a clear difference between the outflows launched by the two binary protostars. In this section, we investigate the origin of the different outflows. The key features we discuss are as follows.

H$_2$ emission originates from the TMC1-E source, while no significant H$_2$ emission from TMC1-W is detected. The H$_2$ emission from TMC1-E has two components: a wind with a narrow opening angle and broader outflow with a wider opening angle, contained within the molecular CO outflow and scattered light emission. Prominent emission from [Fe II] and [Ni II] toward TMC1-W is detected, while both sources show emission from [Ar II] and [Ne II], with the TMC1-E outflow showing bright and radially resolved emission in those tracers.

\begin{figure*}
    \centering
    \includegraphics[width=0.9\textwidth]{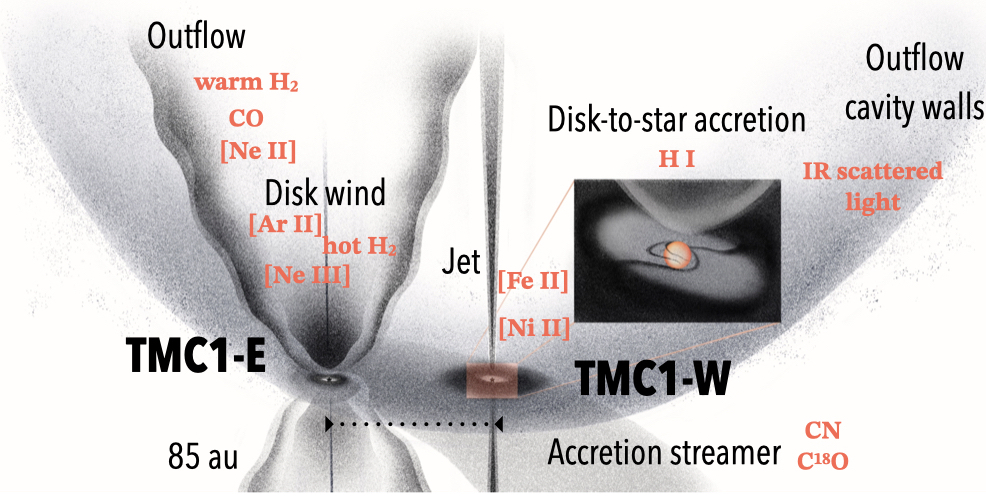}
    \caption{Schematic of physical components and associated tracers for the TMC1 protobinary system.}
    \label{fig:summary}
\end{figure*}

Our proposed scenario that explains the observed features is that different modes of accretion influence the origin of two outflows: the wide-angle H$_2$ wind towards TMC1-E is influenced by the accretion from the large-scale envelope onto the disk, which provides additional material, while the disk-to-star accretion is related to the launch of the powerful, collimated jet from TMC1-W. Figure \ref{fig:summary} shows an illustrative schematic diagram of the observed features. We present below a detailed discussion of the results and the proposed scenarios.

\subsection{TMC1-E: Disk wind possibly powered by infall from the envelope to the disk}

The H$_2$ emission originates in the warm and hot molecular gas in the outflows; however, without spatially resolving the emission, it is difficult to pinpoint the specific location of the hot gas. With MIRI-MRS, we see that the highest \eup\ transitions of H$_2$ are more collimated than the low \eup\ ones. We interpret this as H$_2$ tracing the disk wind, with a flow radius that is larger than the jet, but clearly distinct from the broader entrained envelope material at the outflow cavity walls \ref{fig:h2_integrated_maps}. This is further corroborated by a comparison with the ALMA CO 2--1 map and {\it HST} NICMOS scattered light image (Fig. \ref{fig:ALMA_maps}). Both delineate the extent of the outflow cavity walls, showing that high \eup\ H$_2$ transitions are within this broader component traced by CO, scattered light and lower energy H$_2$ lines. Velocity stratification of the wind and interplay between the jet and the surrounding gas can be responsible for the observed morphology of the gas. In theoretical works, both jet-driven bow shocks \citep[e.g.,][]{Rabenanahary.Cabrit.ea2022a}, and wind-driven cavities can explain those structures \citep[e.g.,][]{Shang.Liu.ea2023}.

With ALMA, we also detected a prominent accretion streamer infalling onto the binary. Tentatively, we associate the TMC1-E with this streamer's infall point, as SO and CN maps suggest. That infall could trigger the prominent disk wind from this source by providing fresh material to replenish the disk. The ionized tracers [Ar II], [Ne II], and [Ne III] are all bright toward this source. Since those noble gas ions have an ionization potential above 13.6 eV, they originate most likely from UV-irradiated material in the disk wind. We base this on the shape of the emission, clearly extended around the base of the wind and not collimated, as seen in the case of [Fe II] or [Ni II] lines. While the disk wind and related UV-excitation is one possibility, some models predict decollimation of the jet in binary systems, leading to a similarly broad appearance of the jet \citep{Lynch.Smith.ea2020}.

The line ratios observed toward the TMC1-E wind position suggest UV-irradiated environments. For example, [Ne III] / [Ne II] = 0.04 is consistent with soft EUV irradiation dominating the excitation and ionization of Ne \citep{Hollenbach.Gorti2009, Szulagyi.Pascucci.ea2012, Espaillat.Ingleby.ea2013}. We detected OH lines towards TMC1-E (Fig. \ref{fig:spectra_overview_outflows}). Those lines are an indicator of UV photodissociation of H$_2$O  \citep[e.g.,][]{Tabone.vanHemert.ea2021}. It is interesting to note that H$_2$O appears more prominent in the gas phase in TMC1-W compared to TMC1-E (Fig. \ref{fig:spectra_overview}); this could be a result of significant water destruction in the TMC1-E source. Ionization and the presence of charged particles can also indicate strong magnetic fields towards this source, which could translate to more prominent disk wind. 

While UV-irradiation is a plausible explanation for the observed ionized species close to the source, further out in the jet, strong, high-velocity shocks ($>$ 100 km s$^{-1}$) can also explain this emission \citep{Hollenbach.McKee1989}. This is especially valid in the case of the TMC1-W jet, where velocities are approaching 200 km s$^{-1}$, stronger emission coming from, for example, [Ar II] is observed further out in the jet in high-velocity channels (see Fig. \ref{fig:chanmaps_v1}).

Interestingly, in the best-studied Class I disk wind source, DG Tau B, the ejection rate in the wind is significantly higher than that of the collimated jet and several times higher than the mass accretion rate on the protostar \citep{Podio.Eisloeffel.ea2011,deValon.Dougados.ea2020}. Our results show that a potential source of this high mass-loss rate in the wind could be the influx of the envelope material. Streamers of gas into the protostar system could therefore have an essential impact on the evolution of the protostellar system. Both observations and simulations suggest they could bring a significant fraction of gas and dust to the protostar and disk -- even at later stages of evolution \citep{ValdiviaMena.Pineda.ea2022, Garufi.Podio.ea2022, Kuffmeier.Jensen.ea2023}.

From a theoretical point of view, simulations of black hole binaries suggest that if the binaries are formed by disk fragmentation, the accretion from the envelope occurs preferentially onto the secondary component, eventually leading to the two components growing similar in mass \citep{Farris.Duffell.ea2014}. \cite{Tokovinin.Moe2020} used those simulation results to successfully reproduce protostellar binaries, explaining the scarcity of brown dwarfs in binary systems. This would be consistent with our ALMA observations indicating the infall of the streamer onto the secondary component. This has been observed in another protostellar system BHB2007 \citep{Alves.Caselli.ea2019}.
On the other hand, in \cite{Murillo.vanDishoeck.ea2022}, the large-scale infall occurs on the more massive component of the IRAS16293 protobinary system. However, in the case of that study, a separation between protostars is 720 au, therefore, it is not likely to have been formed as a result of disk fragmentation \citep{Offner.Kratter.ea2010}. 

What is particularly striking in this system is that the protostellar jets are launched almost perfectly in parallel. From this, we can assume that the disks from which those jets are launched have co-planar geometry, supporting the protostars' common origin. Interestingly, no interaction or disturbance of the jet from TMC1-W by the TMC1-E wind is observed, in contrast to expectations of the models, where jets from binary systems are expected to cause precession or even to become completely attenuated or engulfed in joint flow \citep{Murphy.Lery.ea2008,Lynch.Smith.ea2020}. Most observations of young binary systems also show precession and deflection of jets \citep{Kwon.FernandezLopez.ea2015,Ching.Lai.ea2016, Fridlund.Liseau1998}.Therefore, the TMC1 system presents the most striking example of parallel jets to date. Standing at odds with other dynamically interacting protostars, this indicates that the disks of the two protostars are co-planar. 
%On the other hand, there are known cases of the parallel jets \citep{Fridlund.Liseau1998}, and associated modeling, which shows that magnetic fields can support collimation of the jet in the parallel configuration \citep{Murphy.Lery.ea2008}

\subsection{TMC1-W: Jet possibly powered by the accretion onto the protostar}

We detected six transitions of hydrogen recombination lines toward the TMC1 system. Those lines are expected to trace high temperature and density environments and are usually associated with stellar accretion \citep{Rigliaco.Pascucci.ea2015,Alcala.Manara.ea2017}.
Detections of brighter HI lines toward TMC1-W compared with TMC1-E, even after accounting for extinction can indicate that the TMC1-W source is undergoing more vigorous accretion. A natural consequence of this should be a release of the excess angular momentum close to the accretion streamlines in the inner disk. According to the MHD jet launching paradigm \citep{Matt.Pudritz2005, Romanova.Ustyugova.ea2005} smaller launching radii should correspond to faster ejection and smaller opening angles, leading to the launch of the highly collimated jet, as observed toward TMC1-W.

The accretion rate measured from the HI (7-6) line is $(0.16-14.8) \times 10^{-11}\ \msun\ {\rm yr}^{-1}$. This is a low value compared to the other Class I systems, more typical for Class II systems. \citep{Fiorellino.Tychoniec.ea2022}. For the same HI line in a more massive and younger system, values can be orders of magnitude higher \citep{Beuther.vanDishoeck.ea2023}.

It is possible to have a rough estimate of the ejection rate from the luminosity of the [Fe II] 25.96 $\mu$m line and using Equation 2 in \cite{Watson.Calvet.ea2016}. This latter allows us to derive a result of  [Fe II] 25.96 $\mu$m  $\dot{M}_{wind}$ relation empirically scaling the relation between [O I] 63.2 $\mu$m and $M_{wind}$ provided by  \cite{Hollenbach.McKee1989}. Using the extinction-corrected flux measured towards the total extent of the blueshifted jet of TMC1-W, we obtained a line luminosity 
of  4.8 $\times 10^{-6} {\textrm L}_{\odot}$. We derived a mass-loss rate of 6.7 $\times 10^{-9}\ \msun\ {\rm yr}^{-1}$.

Compared with the accretion rate measured on the source, the ejection rate is 45-4000. It is therefore difficult to reconcile with the models of magnetohydrodynamically-driven winds, where the value ranges from 0.01 to more than 0.5 depending on the details of the model \citep{Nisini.Antoniucci.ea2018}. It is possible that the accretion rate is highly variable and was much higher in the past or that the flux from the [Fe II] line is coming from  jets of both protostars.

Other methods can be also used to extract mass-loss rate from the observed lines, such as measure of the total number of emitting molecules \citep{Dionatos.Nisini.ea2009}. However, the depletion of iron onto the dust grains would contribute to the uncertainty of the measurement in the context of this work.

It is noteworthy that primarily refractory species such as iron, nickel, and chlorine are present in the jet. All those species are expected to be incorporated into the dust grains in the interstellar medium. Therefore, their presence in the gas phase is directly linked to the destruction of the grains. Shock models reveal that only up to 10\% of the refractory material is typically released into the gas phase due to grain destruction in shocks \citep{Gusdorf.Cabrit.ea2008, Gusdorf.PineaudesForets.ea2008}. Observational work to date also confirms that refractory abundances in the jet are depleted with respect to the solar values, indicating that the bulk refractory material is still in the solid state \citep{Podio.Bacciotti.ea2006, Podio.Eisloeffel.ea2011, Dionatos.Nisini.ea2009, Giannini.Antoniucci.ea2015}.

We interpret the presence of refractories as a sign of jet launching within the dust sublimation radius in the protoplanetary disk, where the iron and nickel are released into the gas phase and further launched with the jet \citep[e.g.,][]{Podio.Eisloeffel.ea2011}  This further supports the idea that the TMC1-W source is dominated by the highly collimated jet launched from the inner regions of the disk triggered by strong accretion onto the protostar. The difference between TMC1-E and TMC1-W can also stem from the two jets having different dust content, namely, the more dusty TMC1-E jet has a lower fraction of the refractory material released in the gas phase.

\subsection{Alternative scenarios}

We provide a scenario where the accretion activity in the system is mainly responsible for the striking difference in the outflow behavior of the two protostars. Here, we consider alternative scenarios. It is possible that the two protostars have followed different evolution histories and that, at present, TMC1-W has ended up to be more evolved than TMC1-E. Its lower extinction and stronger ionized jet, without any significant molecular wind component, would support this.

Alternatively, the outflow behavior could be governed by the properties of protoplanetary disks surrounding both protostars. ALMA images show compact dust continuum emission from both targets. If the TMC1-W disk is truncated due to effects such as magnetic braking, this could also make the production of the disk wind difficult. Magnetization of the disk could also play an essential role in the launching of the wind. Therefore, different magnetic field properties in the components could contribute to these differences. It is known that binarity plays an important role in the accretion process, disk truncation, and outbursting activity \citep{Zagaria.Rosotti.ea2021} Lastly, even if the system followed a similar evolution process, it is possible that inflow of the material would cause apparent rejuvenation of the protostar by increasing the amount of material available for accretion \citep{Kuffmeier.Jensen.ea2023}.

\section{Conclusions}
JWST MIRI-MRS observations of Class I protobinary system TMC1 have revealed an unprecedented view of the accretion and ejection processes in protostars. The results from our study are summarized as follows:

\begin{itemize}
    \item We detect two distinct outflows from both protostars, TMC1-W and TMC1-E. The outflow from TMC1-W shows collimated emission lines of [Fe II] and [Ni II] tracing the energetic high-velocity jet. At the same time, TMC1-E powers a prominent wide-angle traced with H$_2$ and is also associated with more collimated emission from [Ne II], [Ne III], and [Ar II]. Jets show a rare parallel geometry that has not been seen before in binary systems, indicating that disks are co-planar.
    \item Six hydrogen recombination lines are detected toward both targets. The lines are much brighter toward TMC1-W than towards TMC1-E, which translates to a higher accretion luminosity and, thus, to higher accretion rates of the TMC1-W source, even when accounting for a higher extinction towards the TMC1-E source. The accretion activity of TMC1-W is likely related to the stronger ejection of material via a collimated jet from this source.
    \item ALMA CO 2--1 map and scattered light emission from HST at 1.6 $\mu$m show that hot H$_2$ traced with MIRI come from within the outflow cavity walls, most likely tracing the disk wind from the TMC1-E source. At the same time, with ALMA, we detect an infalling streamer of gas with CN and SO, pointing to TMC1-E as the landing point of the infall. This could influence the disk growth and transport of the angular momentum from the envelope to the disk and is possibly related to the powerful disk wind launched from this source.
    
\end{itemize}

In this work, we observe a remarkable difference in the appearance of the outflow from this binary system in different atomic and molecular tracers. This comparison allows us to achieve a better understanding of what physical components are probed by different MIR tracers. In this context,  JWST-MIRI has opened up a new avenue of study. Additionally, by linking the outflow properties with stellar accretion measured with our MIR data and infall tracers, as seen by ALMA, we offer a scenario that allows us to explain the differences between outflows that originate in different modes of accretion and infall governing the ejection towards the protostars. This provides a new insight into the connection between accretion and ejection that the models of protostellar collapse and star and disk formation should further investigate. 

%With these observational results of a Class I binary protostar system, we show that accretion and ejection processes in protostars are clearly intertwined, with accretion from the envelope onto the disk triggering a wide-angle wind while accretion from the protostellar disk onto the protostar launching a collimated high-velocity jet.

\begin{acknowledgements} Authors would like to thank the referee for helpful suggestions that improved the clarity of the paper. Ł.T.  would like to thank Rajika Kuruwita, Aashish Gupta, and Ugo Lebreuilly for helpful discussions and Marta Tychoniec for contributing to Fig. 9. This work is based on observations made with the NASA/ESA/CSA James Webb Space Telescope. The data were obtained from the Mikulski Archive for Space Telescopes at the Space Telescope Science Institute, which is operated by the Association of Universities for Research in Astronomy, Inc., under NASA contract NAS 5-03127 for JWST. These observations are associated with program \#1290.

The following National and International Funding Agencies
funded and supported the MIRI development: NASA; ESA; Belgian Science
Policy Office (BELSPO); Centre Nationale d’Etudes Spatiales (CNES); Danish
National Space Centre; Deutsches Zentrum fur Luft- und Raumfahrt (DLR);
Enterprise Ireland; Ministerio De Economiá y Competividad; Netherlands Research
School for Astronomy (NOVA); Netherlands Organisation for Scientific
Research (NWO); Science and Technology Facilities Council; Swiss Space Office;
Swedish National Space Agency; and UK Space Agency. 

H.B. acknowledges support from the Deutsche Forschungsgemeinschaft in the Collaborative Research Center (SFB 881) “The Milky Way System” (subproject B1). EvD, MvG, LF, KS, WR and HL acknowledge support from ERC Advanced grant 101019751 MOLDISK, TOP-1 grant 614.001.751 from the Dutch Research Council (NWO), the Netherlands Research School for Astronomy (NOVA), the Danish National Research Foundation through the Center of Excellence “InterCat” (DNRF150), and DFGgrant 325594231, FOR 2634/2. P.J.K. acknowledges financial support from the Science Foundation Ireland/Irish Research Council Pathway programme under
Grant Number 21/PATH-S/9360. A.C.G. has been supported by PRIN-INAF
MAIN-STREAM 2017 “Protoplanetary disks seen through the eyes of new generation instruments” and from PRIN-INAF 2019 “Spectroscopically tracing the disk dispersal evolution (STRADE)”. K.J. acknowledges the support from the Swedish National Space Agency (SNSA).  G.P. gratefully acknowledges support from the Max Planck Society. This paper makes use of the following ALMA data: ADS/JAO.ALMA\# 2017.1.01350.S, ADS/JAO.ALMA\# 2017.1.01413.S. ALMA is a partnership of ESO (representing its member states), NSF (USA) and NINS (Japan), together with NRC (Canada), MOST and ASIAA (Taiwan), and KASI (Republic of Korea), in co-operation with the Republic of Chile. The Joint ALMA Observatory is operated by ESO, AUI/NRAO and NAOJ. Astrochemistry in Leiden is supported by the Netherlands Research School for Astronomy (NOVA). This research has made use of NASA's Astrophysics Data System Bibliographic Services. This research made use of NumPy \citep{Harris.Millman.ea2020}; Astropy, a community-developed core Python package for Astronomy \citep{ AstropyCollaboration.Robitaille.ea2013,AstropyCollaboration.PriceWhelan.ea2018}; Matplotlib \citep{Hunter2007}; pdrtpy \citep{Kaufman.Wolfire.ea2006,Pound.Wolfire2008,Pound.Wolfire2011,Pound.Wolfire2023}

\end{acknowledgements}

\bibliography{bibjoys.bib}

\appendix

\section{Additional figures}

\subsection{Extinction measurement}
\label{sec: extinction}

\begin{figure*}
    \centering
    \includegraphics[width=0.42\textwidth]{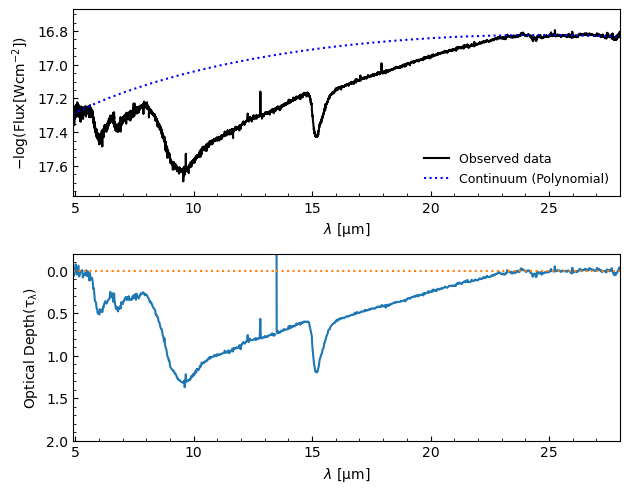}
        \includegraphics[width=0.42\textwidth]{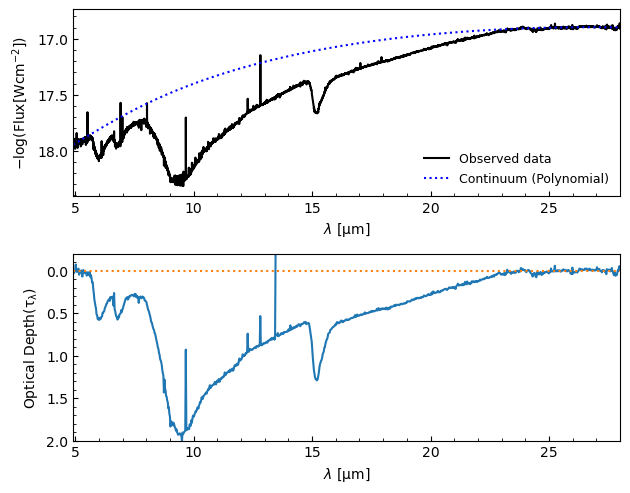}

    \caption{ Spectra extracted from 2 x PSF FWHM on the continuum position (top) of TMC1-W (top-left) and TMC1-E (top-right). The dashed blue line represents a polynomial fit to the continuum. optical depth as a function of wavelength resulting from dividing the polynomial fit with the observed continuum (bottom). The orange dashed line represents $\tau = 0$. The range of extincition is calculated using optical depth value at the silicate feature following the equations: $A_V=\tau_{9.7}\times18.5$, and $A_V=\tau_{9.7}\times9$, representing the correlation found in diffuse ISM and galactic centre dense clouds, respectively \citep{McClure2009, Weingartner.Draine.ea2001, Chiar.Ennico.ea2007, Boogert.Tielens.ea2000}.}
    \label{fig:extcorr_silicate}
    
\end{figure*}

Extinction at the continuum position is measured with the optical depth of the silicate feature at 9.7 $\mu$m. We show the fitting results in Fig. \ref{fig:extcorr_silicate}. Polynomial is fit for the selected wavelength ranges: 4.9--5.3 $\mu$m, 23.8--23.9 $\mu$m, and 25--28 $\mu$m. This allows us to avoid major broad absorption features such as water and silicates.

\begin{figure*}
    \centering
    \includegraphics[width=0.42\textwidth]{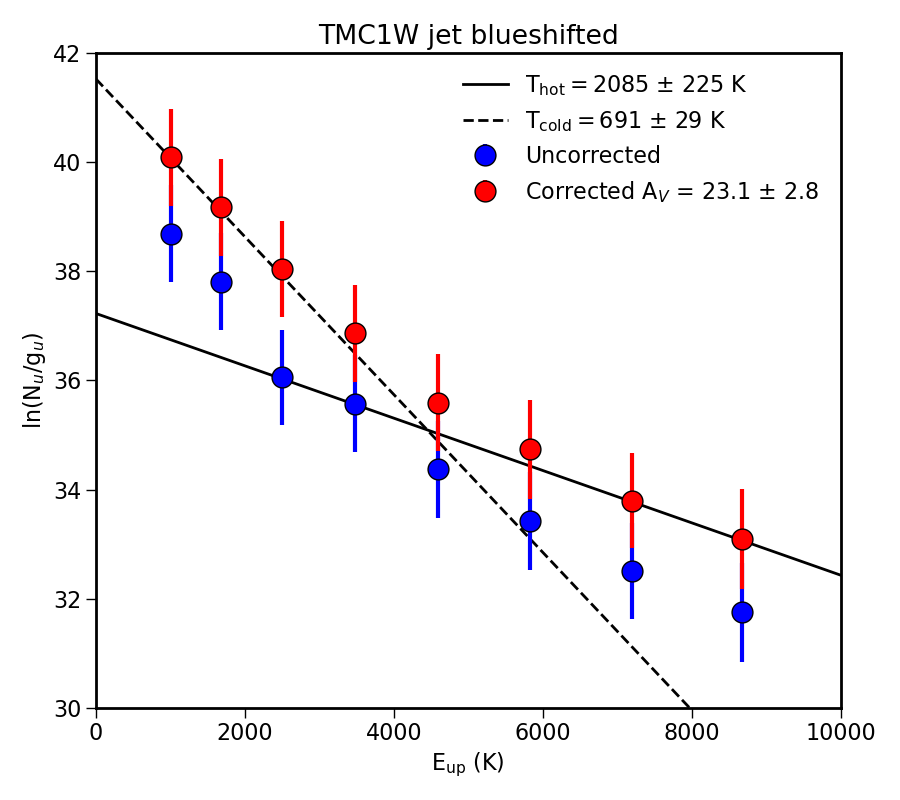}
        \includegraphics[width=0.42\textwidth]{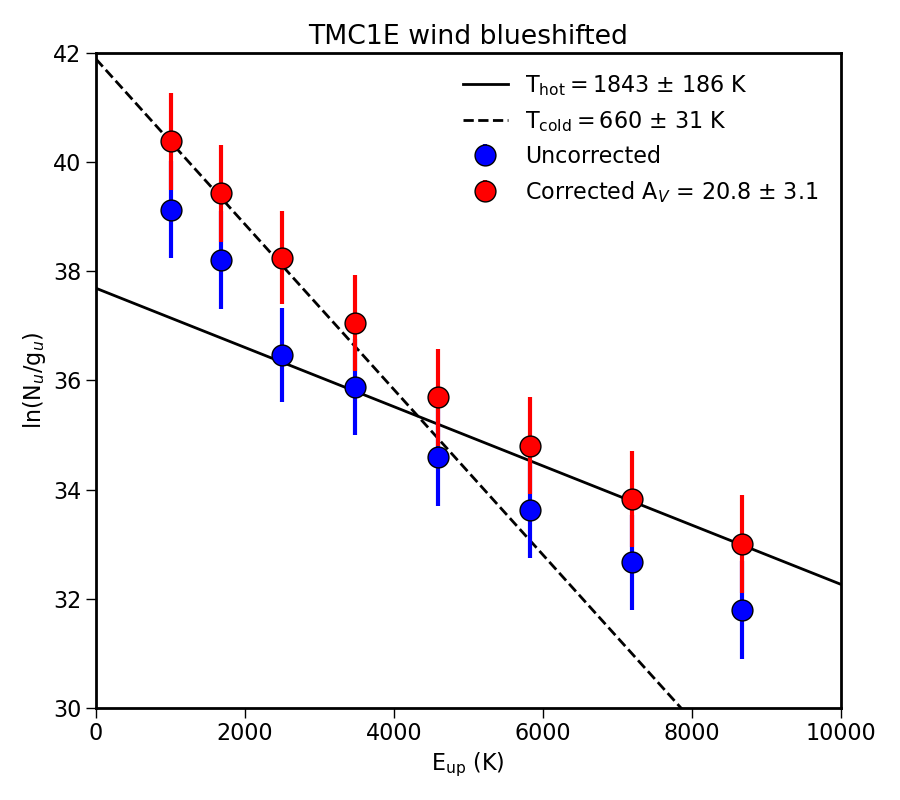}

    \caption{Rotational diagrams for two positions: in the TMC1-W blueshifted jet positions (left) and TMC1-E blueshifted wind positions (right). Regions are indicated in Fig. \ref{fig:representative_regs}.}
    \label{fig:extcorr_h2}
    
\end{figure*}

At the outflow positions, the extinction correction is provided by simultaneously fitting rotational temperature and extinction value.  Fig. \ref{fig:extcorr_h2} shows the results of the fitting. We fit the value of K-band extinction A$_K$.  The relation between A$_\lambda$ and A$_K$ is provided following the extinction curve in \cite{McClure2009} and convert to visual extinction with A$_V$/A$_K$ = 7.75 for R$_V$ = 5.

\begin{figure*}
    \centering
    \includegraphics[width=0.42\textwidth]{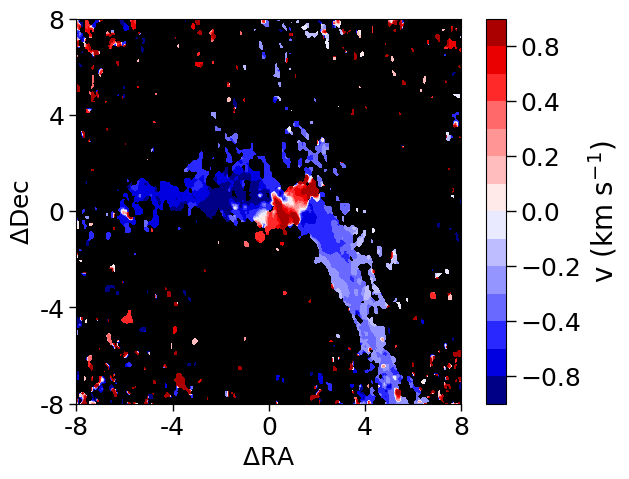}
    \caption{    
    Intensity weighted velocity (moment 1) map of CN (2--1). The map reveals the gradient towards the protostar, consistent with infalling gas. }
    \label{fig:cnmom1}
\end{figure*}

Figure \ref{fig:cnmom1} presents the intensity-weighted channel map of CN, as observed with ALMA. The velocity gradient in the image of increasing velocity with distance to the source is consistent with the infalling streamer.

\begin{figure*}
    \centering
    \includegraphics[width=0.92\textwidth]{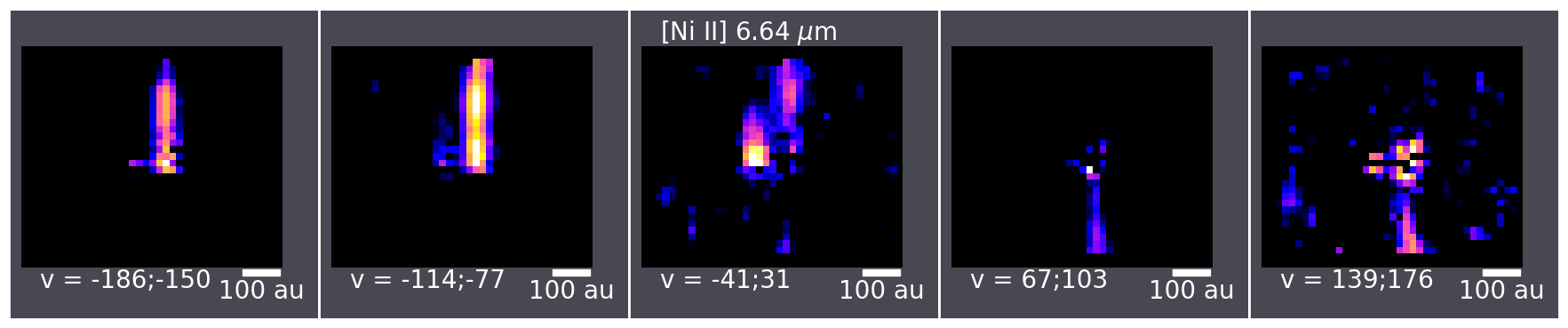}
    \includegraphics[width=0.92\textwidth]{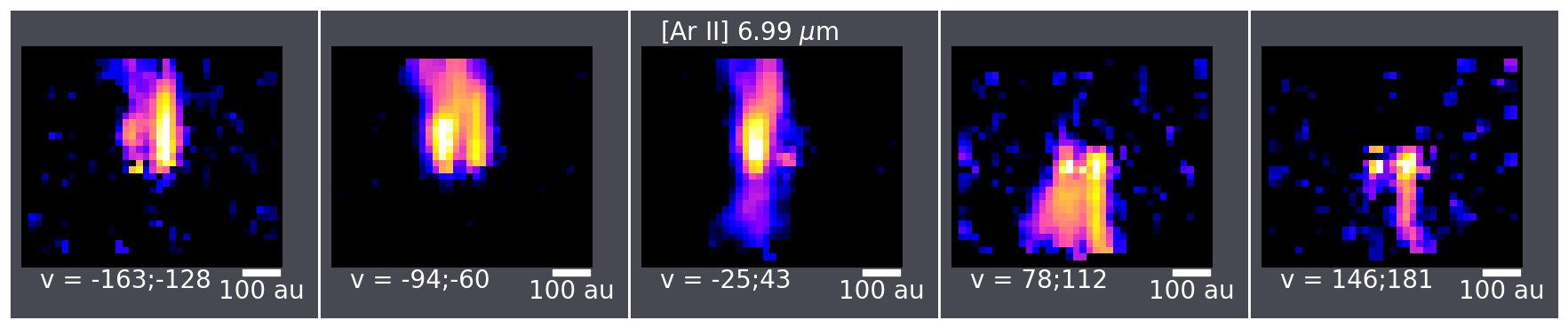}
        \includegraphics[width=0.92\textwidth]{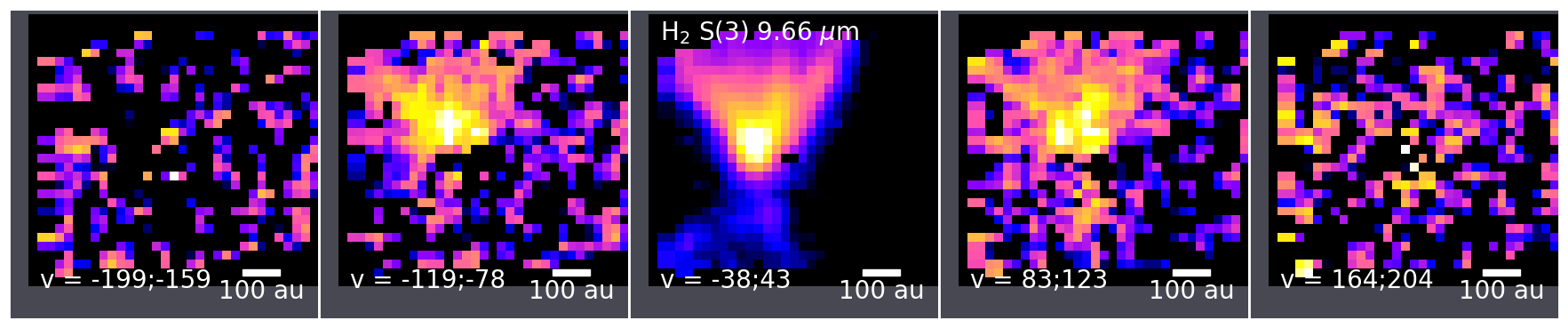}
    \caption{Channel maps for [Ni II],[Ar II] and H$_2$ S(3) showing blueshifted emission toward the north and weaker redshifted emission to the south.
}
    \label{fig:chanmaps_v1}
\end{figure*}

The MRS spectral resolving power varies from R = $\lambda/\Delta \lambda\sim 1500-3500$ \citep{Jones.AlvarezMarquez.ea2023}. In Fig. \ref{fig:chanmaps_v1}, we present the emission from [Ni II], [Ar II], and H$_2$. The emission from the ionized species appears to be velocity-resolved, with clear red- and blueshifted components on the south and north sides of the source, respectively. On the other hand, H$_2$ remains unresolved in the velocity regime, with only faint emission detected above 80 km s$^{-1}$.

\newpage
\section{Tables}

\begin{table*}
\centering
\caption{Line intensities of hydrogen recombination lines at the positions of the protostars.}
\label{tab:hydrogen_both}
\begin{tabular}{lcccccccc}
\hline \hline
Line & $\lambda$ & Intensity & Intensity & Intensity & Intensity \\
 & & uncorrected &  A$_v$ = 18 mag & uncorrected &  A$_v$ = 27 mag  \\
 &  & TMC1-W  & TMC1-W & TMC1-E  & TMC1-E \\
 & $\mu$m &  W m$^{-2}$ arcsec$^{-2}$  &  W m$^{-2}$ arcsec$^{-2}$ & W m$^{-2}$ arcsec$^{-2}$  &  W m$^{-2}$ arcsec$^{-2}$  \\
\hline
    HI(9-6) & 5.91  & $  1.3 \pm \ 0.4 \times 10^{-18} $ & $  3.7 \pm \ 1.2 \times 10^{-18} $ & $  3.4 \pm \ 0.7 \times 10^{-20} $ & $  1.8 \pm \ 0.4 \times 10^{-19} $ \\
    HI(6-5) & 7.46  & $  1.7 \pm \ 0.3 \times 10^{-18} $ & $  4.3 \pm \ 0.8 \times 10^{-18} $ & $  2.5 \pm \ 0.7 \times 10^{-19} $ & $  9.8 \pm \ 2.6 \times 10^{-19} $ \\
    HI(8-6) & 7.50  & $  1.5 \pm \ 0.4 \times 10^{-18} $ & $  3.8 \pm \ 1.1 \times 10^{-18} $ & $  1.5 \pm \ 0.7 \times 10^{-19} $ & $  5.9 \pm \ 2.6 \times 10^{-19} $ \\
    HI(10-7) & 8.76  & $  4.3 \pm \ 1.2 \times 10^{-19} $ & $  1.5 \pm \ 0.4 \times 10^{-18} $ & $  6.4 \pm \ 2.7 \times 10^{-20} $ & $  4.2 \pm \ 1.8 \times 10^{-19} $ \\
    HI(9-7) & 11.31  & $  2.3 \pm \ 0.5 \times 10^{-19} $ & $  7.6 \pm \ 1.5 \times 10^{-19} $ & $  6.8 \pm \ 3.3 \times 10^{-20} $ & $  4.0 \pm \ 2.0 \times 10^{-19} $ \\
    HI(7-6) & 12.37  & $  2.5 \pm \ 0.5 \times 10^{-19} $ & $  7.2 \pm \ 1.4 \times 10^{-19} $ &  $  7.2 \pm \ 1.6 \times 10^{-20} $ & $  3.5 \pm \ 0.8 \times 10^{-19} $ \\
 \hline
\end{tabular}
\end{table*}

\begin{table*}
\centering
\caption{Line intensities of H$_2$ pure rotational transitions integrated at the outflow position of TMC1-E.}
\label{tab:h2_tmc1e}
\begin{tabular}{lcccccccc}
\hline \hline
Line & $\lambda$ & $E_{\rm up}$& Intensity & Intensity \\
 & & & uncorrected & A$_v$ = 20.8 mag \\
 & $\mu$m & K &  W m$^{-2}$ arcsec$^{-2}$ & W m$^{-2}$ arcsec$^{-2}$\\
\hline
H$_2$(1–1)S(9) & 4.95 & 15725  & $  5.2 \pm \ 1.6 \times 10^{-20} $ & $  1.8 \pm \ 0.5 \times 10^{-19} $  \\
H$_2$(1–1)S(7) & 5.81 & 12816  & $  3.9 \pm \ 1.1 \times 10^{-20} $ & $  1.3 \pm \ 0.4 \times 10^{-19} $  \\
H$_2$(1–1)S(5) & 7.28 & 10340  & $  3.2 \pm \ 0.7 \times 10^{-20} $ & $  9.1 \pm \ 1.9 \times 10^{-20} $  \\
H$_2$(0–0)S(8) & 5.05 & 8677  & $  3.2 \pm \ 0.1 \times 10^{-19} $ & $  1.1 \pm \ 0.1 \times 10^{-18} $  \\
H$_2$(0–0)S(7) & 5.51 & 7196  & $  1.2 \pm \ 0.1 \times 10^{-18} $ & $  3.8 \pm \ 0.1 \times 10^{-18} $  \\
H$_2$(0–0)S(6) & 6.11 & 5829  & $  4.7 \pm \ 0.1 \times 10^{-19} $ & $  1.6 \pm \ 0.1 \times 10^{-18} $  \\
H$_2$(0–0)S(5) & 6.91 & 4586  & $  1.5 \pm \ 0.1 \times 10^{-18} $ & $  4.5 \pm \ 0.1 \times 10^{-18} $  \\
H$_2$(0–0)S(4) & 8.03 & 3474  & $  6.1 \pm \ 0.1 \times 10^{-19} $ & $  2.0 \pm \ 0.1 \times 10^{-18} $  \\
H$_2$(0–0)S(3) & 9.66 & 2503  & $  8.5 \pm \ 0.1 \times 10^{-19} $ & $  5.1 \pm \ 0.1 \times 10^{-18} $  \\
H$_2$(0–0)S(2) & 12.28 & 1682  & $  2.9 \pm \ 0.1 \times 10^{-19} $ & $  1.0 \pm \ 0.1 \times 10^{-18} $  \\
H$_2$(0–0)S(1) & 17.03 & 1015  & $  2.1 \pm \ 0.1 \times 10^{-19} $ & $  7.5 \pm \ 0.1 \times 10^{-19} $  \\
\hline
\end{tabular}
\end{table*}

\begin{table*}
\centering
\caption{Line intensities at the outflow positions}
\label{tab:ions_both}
\begin{tabular}{lcccccccc}
\hline \hline
Line & $\lambda$ & Intensity & Intensity & Intensity & Intensity \\
 & & uncorrected &  A$_v$ = 23.1 mag & uncorrected &  A$_v$ = 20.8 mag  \\
 &  & TMC1-W  & TMC1-W & TMC1-E  & TMC1-E \\
 \hline    
    $[$FeII$]$ $^4$F$_\frac{9}{2}$-$^6$D$_\frac{9}{2}$ & 5.34  & $  9.3 \pm \ 0.3 \times 10^{-19} $ & $  3.5 \pm \ 0.1 \times 10^{-18} $ & $  1.0 \pm \ 0.2 \times 10^{-19} $ & $  3.4 \pm \ 0.5 \times 10^{-19} $ \\
    $[$NiII$]$ $^2$D$_\frac{5}{2}$-$^2$D$_\frac{3}{2}$ & 6.64  & $  8.3 \pm \ 0.2 \times 10^{-19} $ & $  2.9 \pm \ 0.1 \times 10^{-18} $ & $  7.1 \pm \ 1.3 \times 10^{-20} $ & $  2.1 \pm \ 0.4 \times 10^{-19} $ \\
    $[$FeII$]$ $^4$F$_\frac{9}{2}$-$^6$D$_\frac{7}{2}$ & 6.72  & $  3.3 \pm \ 0.5 \times 10^{-20} $ & $  1.2 \pm \ 0.2 \times 10^{-19} $ &  $< 3.6 \times 10^{-19}$ & $< 1.1 \times 10^{-18}$ \\
    $[$ArII$]$ $^2$P$_\frac{3}{2}$-$^2$P$_\frac{1}{2}$ & 6.99  & $  5.1 \pm \ 0.1 \times 10^{-19} $ & $  1.8 \pm \ 0.0 \times 10^{-18} $ & $  7.4 \pm \ 2.1 \times 10^{-21} $ & $  3.6 \pm \ 1.0 \times 10^{-20} $ \\
    $[$ArIII$]$ $^3$P$_2$-$^3$P$_1$ & 8.99  & $  1.4 \pm \ 0.6 \times 10^{-20} $ & $  8.0 \pm \ 3.7 \times 10^{-20} $ & $  4.6 \pm \ 0.1 \times 10^{-19} $ & $  1.4 \pm \ 0.0 \times 10^{-18} $ \\
    $[$CoII$]$ $^3$F$_4$-$^3$F$_3$ & 10.52  & $  1.9 \pm \ 0.3 \times 10^{-20} $ & $  1.1 \pm \ 0.2 \times 10^{-19} $ & $  4.4 \pm \ 3.2 \times 10^{-21} $ & $  2.1 \pm \ 1.5 \times 10^{-20} $ \\
    $[$NiII$]$ $^4$F$_\frac{9}{2}$-$^4$D$_\frac{7}{2}$ & 10.68  & $  1.1 \pm \ 0.0 \times 10^{-19} $ & $  5.9 \pm \ 0.2 \times 10^{-19} $ & $  9.5 \pm \ 4.4 \times 10^{-21} $ & $  4.4 \pm \ 2.0 \times 10^{-20} $ \\
    $[$ClI$]$ $^2$P$_\frac{3}{2}$-$^2$P$_\frac{1}{2}$ & 11.33  & $  1.6 \pm \ 0.2 \times 10^{-20} $ & $  7.3 \pm \ 1.1 \times 10^{-20} $ & $  2.2 \pm \ 0.4 \times 10^{-20} $ & $  8.8 \pm \ 1.7 \times 10^{-20} $ \\
    HCO${^+}$ $v=2$ & 12.07  & $  1.1 \pm \ 0.4 \times 10^{-20} $ & $  4.3 \pm \ 1.6 \times 10^{-20} $ & $  3.7 \pm \ 0.7 \times 10^{-20} $ & $  1.3 \pm \ 0.2 \times 10^{-19} $ \\
    $[$NiII$]$ $^4$F$_\frac{7}{2}$-$^4$F$_\frac{5}{2}$ & 12.73  & $  3.0 \pm \ 0.4 \times 10^{-20} $ & $  1.1 \pm \ 0.1 \times 10^{-19} $ & $  1.2 \pm \ 0.8 \times 10^{-20} $ & $  3.8 \pm \ 2.7 \times 10^{-20} $ \\
    $[$NeII$]$ $^2$P$_\frac{3}{2}$-$^2$P$_\frac{1}{2}$ & 12.81  & $  1.8 \pm \ 0.0 \times 10^{-18} $ & $  6.6 \pm \ 0.2 \times 10^{-18} $ & $  2.1 \pm \ 0.0 \times 10^{-18} $ & $  6.9 \pm \ 0.0 \times 10^{-18} $ \\
    HCN $v=2$  & 14.00 & $< 1.2 \times 10^{-19}$ & $< 4.3 \times 10^{-19}$ & $< 1.1 \times 10^{-19}$ & $< 3.6 \times 10^{-19}$ \\
    $[$ClII$]$ $^3$P$_2$-$^3$P$_1$ & 14.37  & $  1.2 \pm \ 0.3 \times 10^{-20} $ & $  4.3 \pm \ 1.2 \times 10^{-20} $ & $  1.3 \pm \ 0.2 \times 10^{-20} $ & $  4.1 \pm \ 0.8 \times 10^{-20} $ \\
    CO$_2$ $v=2$ & 14.97  & $  1.6 \pm \ 0.3 \times 10^{-21} $ & $  6.3 \pm \ 1.3 \times 10^{-21} $ & $  8.0 \pm \ 0.8 \times 10^{-21} $ & $  2.7 \pm \ 0.3 \times 10^{-20} $ \\
    $[$NeIII$]$ $^3$P$_2$-$^3$P$_1$ & 15.56  & $  3.5 \pm \ 0.4 \times 10^{-20} $ & $  1.4 \pm \ 0.2 \times 10^{-19} $ & $  7.0 \pm \ 0.4 \times 10^{-20} $ & $  2.5 \pm \ 0.1 \times 10^{-19} $ \\
    HD (0-0)R(6) & 16.89  & $  4.0 \pm \ 2.0 \times 10^{-21} $ & $  1.6 \pm \ 0.8 \times 10^{-20} $ & $  3.9 \pm \ 2.8 \times 10^{-21} $ & $  1.4 \pm \ 1.0 \times 10^{-20} $ \\
    $[$FeII$]$ $^4$F$_\frac{7}{2}$-$^4$F$_\frac{9}{2}$ & 17.94  & $  1.4 \pm \ 0.0 \times 10^{-18} $ & $  5.7 \pm \ 0.1 \times 10^{-18} $ & $  2.7 \pm \ 0.1 \times 10^{-19} $ & $  9.2 \pm \ 0.2 \times 10^{-19} $ \\
    $[$SIII$]$ $^3$P$_2$-$^3$P$_1$ & 18.71  & $  1.8 \pm \ 0.3 \times 10^{-20} $ & $  7.0 \pm \ 1.3 \times 10^{-20} $ & $  1.8 \pm \ 1.6 \times 10^{-20} $ & $  6.0 \pm \ 5.3 \times 10^{-20} $ \\
    $[$FeII$]$ $^4$F$_\frac{5}{2}$-$^4$F$\frac{7}{2}$ & 24.52  & $  2.7 \pm \ 0.2 \times 10^{-19} $ & $  7.6 \pm \ 0.5 \times 10^{-19} $ & $  9.5 \pm \ 1.9 \times 10^{-20} $ & $  2.4 \pm \ 0.5 \times 10^{-19} $ \\
    $[$SI$]$ $^3$P$_2$-$^3$P$_1$ & 25.25  & $  1.0 \pm \ 0.2 \times 10^{-19} $ & $  2.8 \pm \ 0.4 \times 10^{-19} $ & $  1.7 \pm \ 0.3 \times 10^{-19} $ & $  4.1 \pm \ 0.7 \times 10^{-19} $ \\
    $[$FeII$]$ $^6$D$_\frac{7}{2}$-$^6$D$_\frac{9}{2}$ & 25.99  & $  9.7 \pm \ 0.2 \times 10^{-19} $ & $  2.5 \pm \ 0.1 \times 10^{-18} $ & $  2.4 \pm \ 0.2 \times 10^{-19} $ & $  5.7 \pm \ 0.5 \times 10^{-19} $ \\
    \hline
    \end{tabular}
    \end{table*}

\begin{table*}
\centering
\caption{Extinction-corrected line intensity ratios for selected lines}
\label{tab:line_rations}
\begin{tabular}{lcccccccc}
\hline \hline
Lines & TMC1E & TMC1W  \\
&  &  & & \\
\hline
$[$Fe II$]$ 5.34 / $[$Fe II$]$ 25.99 & 0.60 & 1.36\\
$[$Fe II$]$ 5.34 / $[$Fe II$]$ 17.94 & 0.37 & 0.61\\
$[$Fe II$]$ 17.94 / $[$Fe II$]$ 24.52 & 3.84 & 7.52\\
$[$Fe II$]$ 17.94 / $[$Fe II$]$ 25.99 & 1.61 & 2.24\\
$[$Fe II$]$ 24.52 / $[$Fe II$]$ 25.99 & 0.42 & 0.30\\
$[$Ar II$]$ 6.99 / $[$Ar III$]$ 8.99 & 38.94 & 22.36\\
$[$Ne II$]$ 12.81 / $[$Ne III$]$ 15.56 & 27.72 & 46.06\\
$[$Ne II$]$ 12.81 / $[$Fe II$]$ 17.94 & 7.52 & 1.16\\
$[$Ne II$]$ 12.81 / $[$Fe II$]$ 25.99 & 12.07 & 2.60\\
$[$Ne II$]$ 12.81 / $[$S I$]$ 25.25 & 16.92 & 23.71\\
$[$Ne II$]$ 12.81 / $[$Ar II$]$ 6.99 & 4.93 & 3.71\\
$[$Cl II$]$ 14.37 / $[$Cl I$]$ 11.33 & 0.47 & 0.59\\

\hline
\end{tabular}
\end{table*}

\end{document}